\documentclass{aa}  

\usepackage{graphicx}
\usepackage{txfonts}
\usepackage{cprotect}
\usepackage{gensymb} 
\usepackage{xspace}
\usepackage{caption}
\usepackage{subcaption}
\def\ms{\hbox{m\,s$^{-1}$}}         
\def\cms{\hbox{\,cm\,s$^{-1}$}}       
\def\m2s2{\hbox{\,m$^{2}$\,s$^{-2}$}} 
\def\kms{\hbox{\,km\,s$^{-1}$}}       
\def\vsini{\hbox{$v$\,sin\,$i_{\star}$}}      
\def\Msun{\hbox{$M_{\odot}$}}             
\def\Rsun{\hbox{$R_{\odot}$}}
\def\Mjup{\hbox{$\mathrm{M}_{\rm J}$}}
\def\Rjup{\hbox{$\mathrm{R}_{\rm J}$}}
\def\Mearth{\hbox{$\mathrm{M}_{\rm E}$}}

\def\Kepler{{\it Kepler}}
\def\ten[#1]{$\;\times 10^{#1}$}

\def\logg{$\log g$}
\def\Kepler{{\it Kepler}}

\newcommand{\e}[1]{{\times10^{#1}}}

\newcommand{\Rnom}{\hbox{$\mathcal{R}^{\rm N}_{\odot}$}} 

\newcommand{\GMnom}{\hbox{$\mathcal{(GM)}^{\rm N}_{\odot}$}}

\newcommand{\RJnom}{\hbox{$\mathcal{R}^{\rm N}_{e \rm J}$}}

\newcommand{\GMJnom}{\hbox{$\mathcal{(GM)}^{\rm N}_{\rm J}$}}

\newcommand{\reb}{{\sc \tt REBOUND}\xspace}
\newcommand{\whf}{{\sc \tt WHFast}\xspace}
\newcommand{\emcee}{{\sc \tt emcee}\xspace}

\begin{document} 

\title{SOPHIE velocimetry of \Kepler\ transit candidates}

    \subtitle{XVIII. Radial velocity confirmation, absolute masses and radii, and origin of the Kepler-419 multiplanetary system\thanks{Based on observations made with SOPHIE on the 1.93m telescope at the Observatoire de Haute-Provence (CNRS), France.}\fnmsep\thanks{Table~\ref{table.RV} are available in electronic form at the CDS via anonymous ftp to cdsarc.u-strasbg.fr (130.79.128.5) or via http://cdsweb.u-strasbg.fr/cgi-bin/qcat?J/A+A/}}

   \author{J.M.~Almenara\inst{\ref{geneva}}
     \and R.F.~D\'{i}az\inst{\ref{geneva},\ref{uba},\ref{iafe}}
     \and G.~H\'ebrard\inst{\ref{iap},\ref{ohp}}
     \and R.~Mardling\inst{\ref{australia},\ref{geneva}}     
     \and C.~Damiani\inst{\ref{ias}}
     \and A.~Santerne\inst{\ref{lam}}
     \and F.~Bouchy\inst{\ref{geneva}}
     \and S.C.C.~Barros\inst{\ref{porto}}
     \and I.~Boisse\inst{\ref{lam}}
     \and X.~Bonfils\inst{\ref{grenoble}}   
     \and A.S.~Bonomo\inst{\ref{inaf}}
     \and B.~Courcol\inst{\ref{lam}}
     \and O.~Demangeon\inst{\ref{lam}}
     \and M.~Deleuil\inst{\ref{lam}}
     \and J.~Rey\inst{\ref{geneva}}
     \and S.~Udry\inst{\ref{geneva}}
     \and P.A.~Wilson\inst{\ref{leiden},\ref{iap}}
    }

   \institute{
   Observatoire de Gen\`eve, D\'epartement d’Astronomie, Universit\'e de Gen\`eve, Chemin des Maillettes 51, 1290 Versoix, Switzerland\label{geneva}
    \and Universidad de Buenos Aires, Facultad de Ciencias Exactas y Naturales. Buenos Aires, Argentina.\label{uba}
    \and CONICET - Universidad de Buenos Aires. Instituto de Astronom\'ia y F\'isica del Espacio (IAFE). Buenos Aires, Argentina.\label{iafe}
     \and Institut d'Astrophysique de Paris, UMR7095 CNRS, Universit\'e Pierre \& Marie Curie, 98bis boulevard Arago, 75014 Paris, France\label{iap}
     \and Observatoire de Haute Provence, 04670 Saint Michel l'Observatoire, France\label{ohp}
     \and School of Physics \& Astronomy, Monash University, Victoria, 3800, Australia\label{australia}
     \and Universit\'e Paris-Sud, CNRS, Institut d'Astrophysique Spatiale, UMR8617, 91405 Orsay Cedex, France\label{ias}
     \and Aix Marseille Univ, CNRS, LAM, Laboratoire d'Astrophysique de Marseille, Marseille, France\label{lam}
     \and Instituto de Astrof\'isica e Ci\^{e}ncias do Espa\c co, Universidade do Porto, CAUP, Rua das Estrelas, 4150-762 Porto, Portugal\label{porto}
     \and Univ. Grenoble Alpes, CNRS, IPAG, 38000 Grenoble, France\label{grenoble}
     \and INAF - Osservatorio Astrofisico di Torino, via Osservatorio 20, 10025 Pino Torinese, Italy \label{inaf}
     \and Leiden Observatory, Leiden University, Postbus 9513, 2300 RA Leiden, The Netherlands \label{leiden}
   }

   \date{}

 
  \abstract
      { 

Kepler-419 is a planetary system discovered by the \Kepler\ photometry which is known to harbour two massive giant planets: an inner 3~\Mjup\ transiting planet with a  69.8-day period,  highly eccentric orbit, and an outer 7.5~\Mjup\ non-transiting planet predicted from the transit-timing variations (TTVs) of the inner planet~b to have a 675-day period, moderately eccentric orbit. Here we present new radial velocity (RV) measurements secured over more than two years with the SOPHIE spectrograph, where both planets are clearly detected.
The RV data is modelled  together with the \Kepler\ photometry using a photodynamical model. 
The inclusion of velocity information breaks the $MR^{-3}$ degeneracy inherent in timing data alone, allowing us to measure the absolute stellar and planetary radii and masses.
With uncertainties of 12\% and 13\% for the stellar and inner planet radii, and 35\%, 24\%, and 35\% for the masses of the star, planet~b, and planet~c respectively, these measurements are the most precise to date for a single host star system using this technique. The transiting planet mass is determined at better precision than the star mass. This shows that modelling the radial velocities and the light curve together in systems of dynamically interacting planets provides a way of characterising both the star and the planets without being limited by knowledge of the star.
On the other hand, the period ratio and eccentricities place the Kepler-419 system in a sweet spot; had around twice as many transits been
observed,  the mass of the transiting planet could have been measured using its own TTVs. 
Finally, the origin of the Kepler-419 system is discussed. 
We show that the system is near a coplanar high-eccentricity secular fixed point, related to the alignment  of the orbits, which has prevented the inner orbit from circularising. For most other relative apsidal orientations, planet~b's orbit would be circular with a semi-major axis of 0.03~au. This suggests a mechanism for forming hot Jupiters in multiplanetary systems without the need of high mutual inclinations.

      }
   \authorrunning{J.M. Almenara et al.}
   \titlerunning{}

   \maketitle
%

\section{Introduction}

Kepler-419 (KOI-1474) is a planetary system with two known giant planets. The inner one, Kepler-419b, was first discovered transiting in the \Kepler\ photometry by \citet{borucki2011} with a period of 69.7~days and an estimate radius of 1.0 \Rjup. Its size and relatively long period places it in the `Period Valley' of giant planets \citep{udry2003,batygin2016}. Strong transit-timing variations (TTVs) of the order of an hour were detected later by \citet{ford2012} and \citet{dawson2012}. \citet{dawson2012} concluded that the observed TTVs were consistent with perturbations from a massive, eccentric outer companion in the system, but they could not constrain the outer body’s orbital period or mass because of the small number of transits and poor orbit coverage \Kepler\ data had at the time. Additionally, they validated the planetary nature of the transits, determined through the  photoeccentric effect, that the orbit of the transiting planet is highly eccentric ($e=0.81^{+0.10}_{-0.07}$), and they found that the host star is a rapidly rotating F7 star with a rotational period of P$_{\rm rot}=4.6\pm0.4$~days. Some years later, using 11 quarters of \Kepler\ photometry, \citet{mazeh2015} measured a rotational period of P$_{\rm rot}=4.53\pm0.16$~days.

In 2012 we started a radial velocity follow-up of Kepler-419 with the SOPHIE spectrograph in order to detect and characterise  the outer companion and to refine the parameters of the inner, transiting planet. This target is part of the sample presented in \citet{santerne2016}. During our follow up campaign, \citet{dawson2014} performed a TTV analysis of 16 quarters of \Kepler\ data (containing all transits observed by \Kepler), which allowed them to ascertain that the perturber object is a non-transiting planet, namely Kepler-419c, with a mass of $7.3\pm0.4$~\Mjup\ on an eccentric orbit ($e=0.184\pm0.002$) with a period of $675.47\pm0.11$~days. They also presented 20 radial velocities secured with the HIRES spectrograph, allowing the mass of Kepler-419b to be measured at $2.5\pm0.3$~\Mjup\ and confirming the   photometrically determined eccentricity. The HIRES data also showed an additional acceleration consistent with the Kepler-419c properties derived from TTVs, but were not able to detect the exterior planet independently.

The system is remarkable because it harbours an almost coplanar pair of giants for which the innermost planet has an extremely high eccentricity. Since mechanisms usually invoked to explain highly eccentric orbits require the presence of a companion on a significantly inclined orbit, Kepler-419 presents a challenge to theory to explain its origin. Moreover, the apsidal lines are close to anti-aligned, the ascending node longitudes are close to aligned, and integrations confirm that this configuration persists over  secular timescales. The state of the system is highly suggestive of gentle relaxation at some point in its history, with the source of dissipation coming from either the protoplanetary disk or tides in planet~b (or both). However, the evolutionary path to its current state is still not clear.

Here we present our SOPHIE radial velocity measurements, comprising 45 epochs over 2.2~years. Both planets are detected independently in the new  data set, confirming the TTV detection of the exterior planet. In order to explore the contribution of each data set, we analyse the SOPHIE radial velocities with a simple two-Keplerian model and the \Kepler\ photometry independently of the SOPHIE data. 

On the other hand, and as described in detail by \citet{agol2005} and \citet{almenara2015, almenara2016}, photometry alone makes it possible to measure the density of the bodies of the system, that is $MR^{-3}$. However, individual masses and radii cannot be constrained, and we only have access to mass ratios and radius ratios. In other words, exactly the same light curve is obtained if lengths in the system are scaled by a factor and masses are scaled by the same factor at cubic exponent. This is called the $MR^{-3}$ degeneracy. This degeneracy can be removed by constraining the system scale, for example by adding radial velocities, or by measuring the light-travel time, which provides access to absolute masses and radii. Then we combine all available data and use a photodynamical model \citep{carter2011} to derive absolute physical parameters without theoretical stellar models \citep{agol2005}. Finally, the results from the photodynamical modelling are used to study the evolution of the system's orbital parameters through numerical integrations over 10 kyr.

\section{Data}

\subsection{\Kepler\ light curve}

\begin{figure*}[t]
\hspace{-0.5cm}\includegraphics[width=1.0\textwidth]{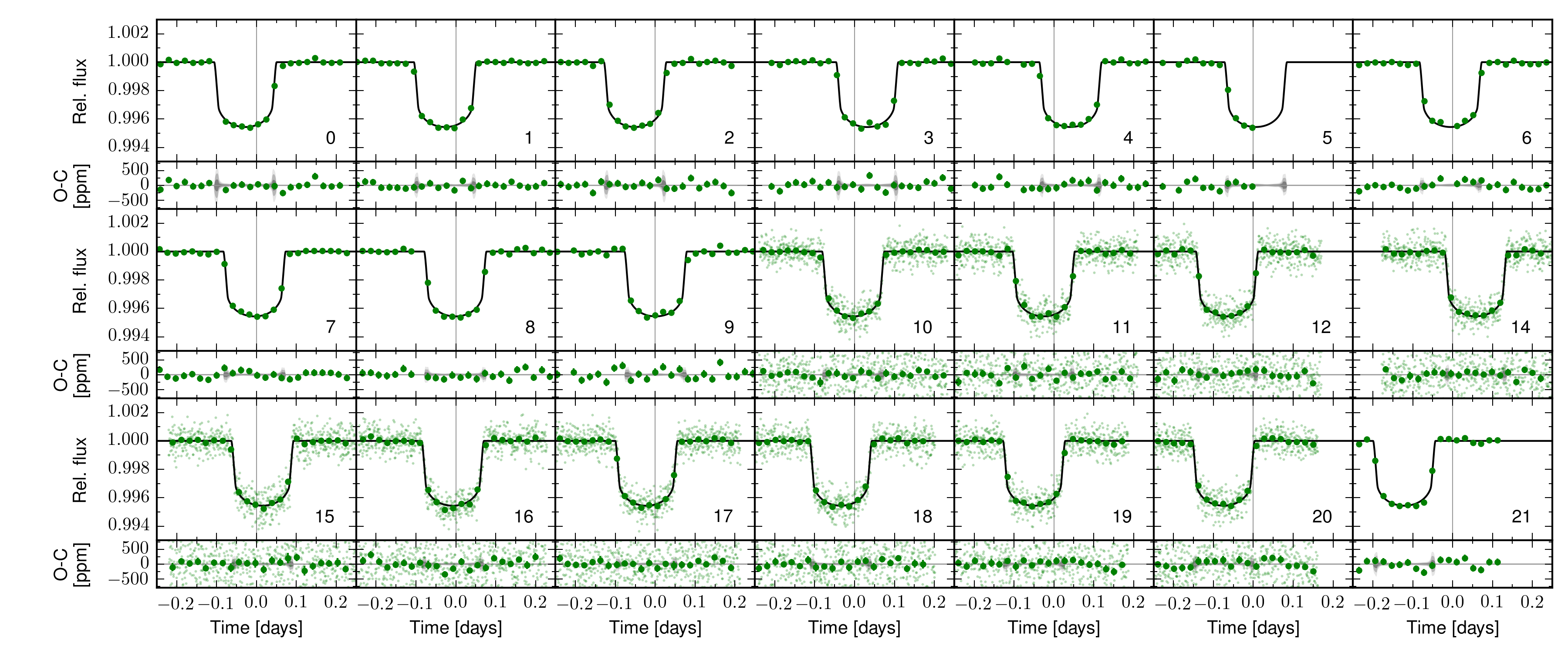}
\caption{Transits of Kepler-419b observed by \Kepler. Each panel is centred at the linear ephemeris (indicated by the vertical grey lines, and reported in the caption of Fig.~\ref{fig.TTV}). For short-cadence data, 29.4-minute binned data is shown in addition to the observed data points. Each panel is labelled with the epoch;  zero is the first transit after $t_{\mathrm{ref}}$. The black curve is the median oversampled model over 10\;000 random MCMC steps. In the lower part of each panel the residuals after subtracting the MAP model to the observed data are shown. The shades of grey represent the 68.3, 95.5, and 99.7\% credible intervals, which are hardly distinguishable in the residual panels and show an increased uncertainty at ingress and egress times.}
\label{fig.PH}
\end{figure*}

\Kepler\ observed the 13-magnitude star Kepler-419 from  Q0 to Q17. We used  Data Release 25 obtained from the Mikulski Archive for Space Telescopes (MAST) archive\footnote{http://archive.stsci.edu/index.html.}. \Kepler\ recorded 21 of the 22 transits of planet~b that took place in the time span of its observation, the first ten and the last one in long-cadence data (about one point every 29.4~min) and the remaining ten in short-cadence data (about one point per minute). We used the simple aperture photometry (SAP) light curve, which we corrected for the flux contamination (between 0 and 2\% depending on the quarter) using the value estimated by the \Kepler\ team. We kept only the data spanning three transit durations around each transit; they were modelled after normalisation using a linear function for each transit. The transit observations are presented in Fig.~\ref{fig.PH}.

\subsection{Radial velocities}

We observed the star Kepler-419 with SOPHIE \citep{perruchot2008,bouchy2009} at the 1.93 m telescope of the Observatoire de Haute-Provence (France). Observations were secured in the slow readout mode of the detector and in high-efficiency mode of the spectrograph, with a resolution power of $\lambda/\Delta\lambda=39\,000$. The first optical fibre was used for starlight, whereas the second fibre was placed on the sky to evaluate the sky background pollution, especially from moonlight. Wavelength calibrations were secured approximately every  two hours during the night to monitor and correct for the potential spectrograph drifts. Forty-five exposures of Kepler-419 were secured between May 2012 and July 2014. Most of them have one-hour exposure time and their signal-to-noise ratios per pixel at 550~nm range from 16 to 43, with a typical value of 30. This translates into a mean radial velocity precision of 28 \ms, estimated following \citet{boisse2010}.

\begin{figure*}
\begin{center}
\begin{minipage}{1.0\textwidth}
\hspace{-0cm}\includegraphics[width=1.0\textwidth]{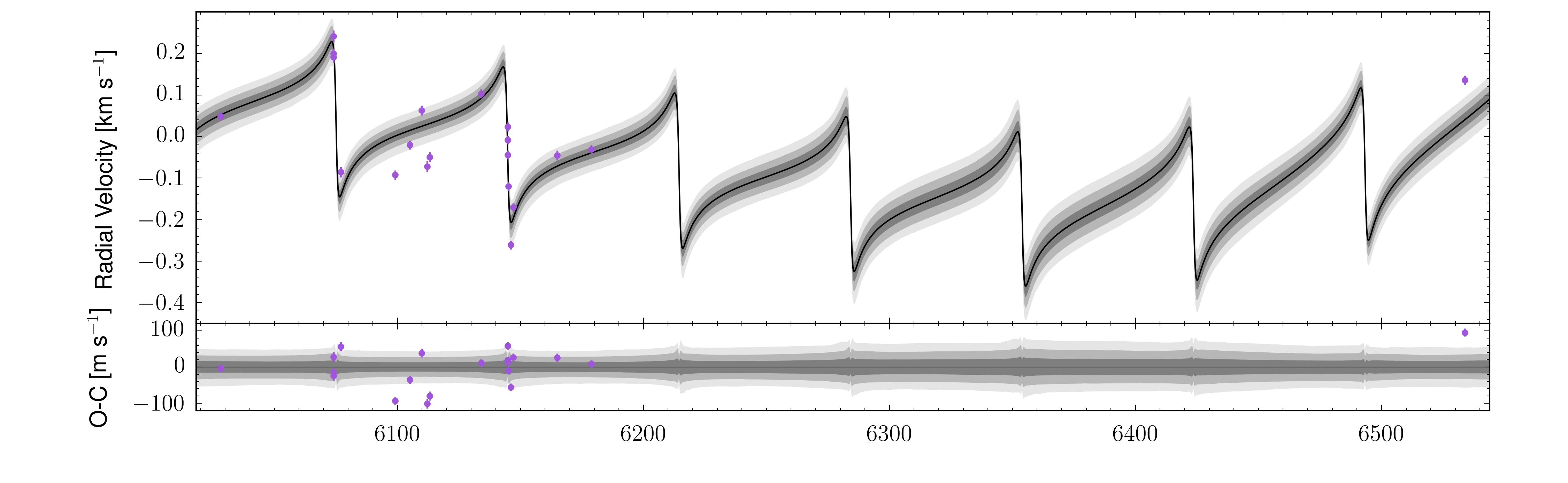}
\end{minipage}
\begin{minipage}{1.0\textwidth}
\hspace{-0cm}\includegraphics[width=1.0\textwidth]{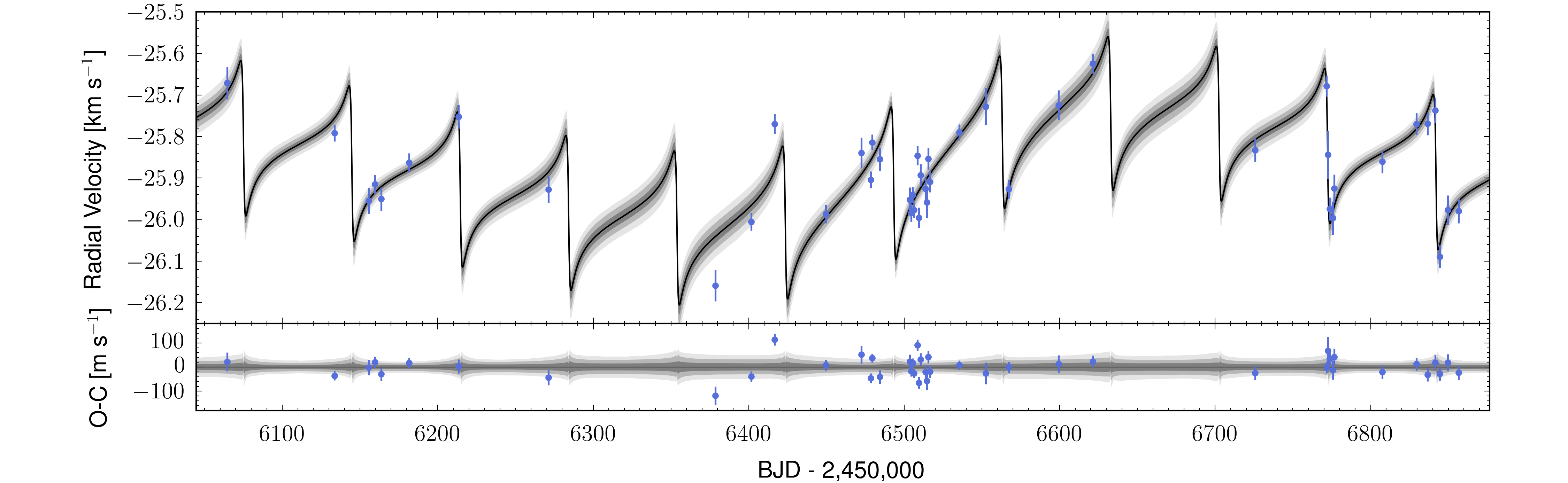}
\end{minipage}
\end{center}
\caption{Radial velocities of Kepler-419 observed by HIRES (upper plot, purple points) and SOPHIE (lower plot, blue points). The black curve is the median model from the photodynamical analysis, with the 68.3, 95.5, and 99.7\% credible curves.}
\label{fig.RV}
\end{figure*}

The spectra were extracted using the SOPHIE pipeline \citep{bouchy2009}, and cross-correlated with a G2-type numerical mask to produce cross-correlation functions (CCFs). CCFs are fit with Gaussians to derive the radial velocities \citep{baranne1996,pepe2002}. The cross-correlation with other types of numerical masks (e.g. K0 ou K5) does not significantly change the results. A series of corrections were subsequently applied, following \citet{santerne2016}:  CCD charge transfer inefficiency was corrected (as exposures have different signal-to-noise ratios), background pollution was removed (which produced corrections of up to 40~\ms\ on ten affected exposures), and instrumental drifts were subtracted using the monitoring of the constant star HD\,185144 (which shows a dispersion of 6.5~\ms\ over our 2.2-year time span). The resulting radial velocities are listed in Table~\ref{table.RV} and plotted in Fig.~\ref{fig.RV}.

\begin{figure}
\includegraphics[width=0.48\textwidth]{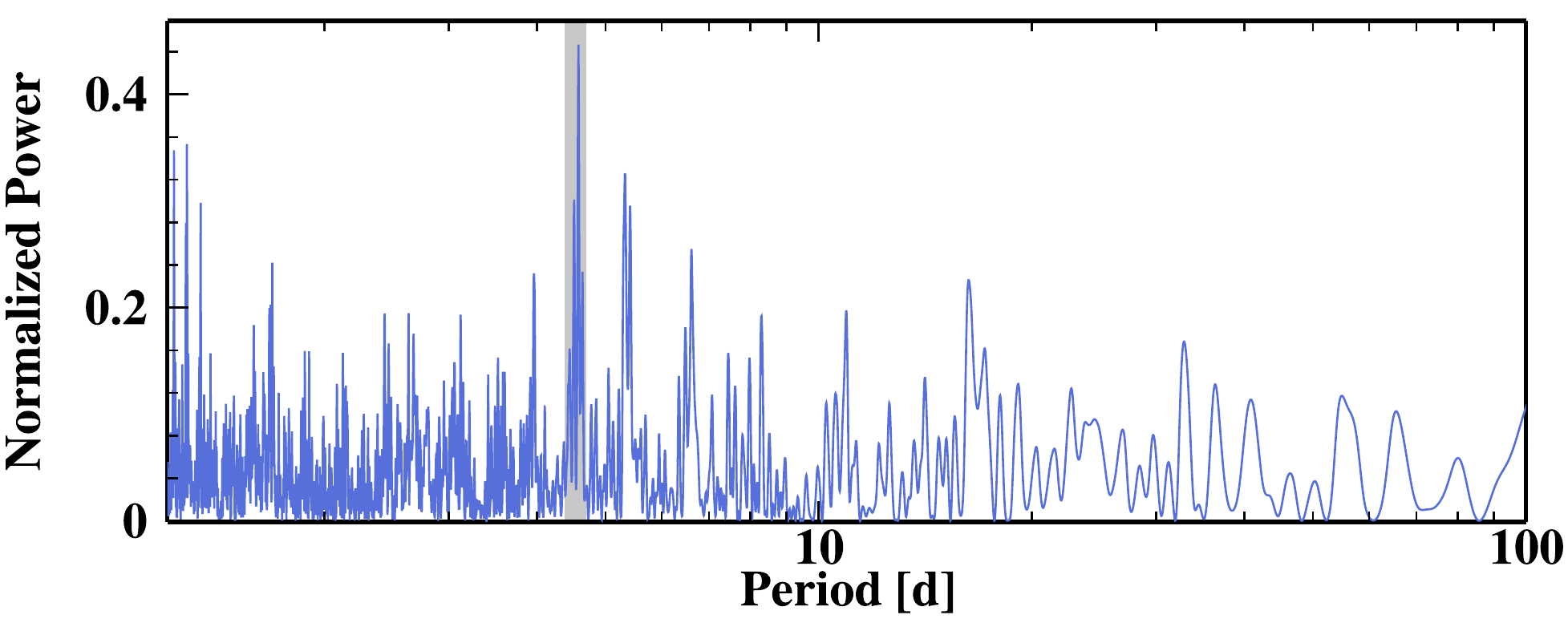}
\caption{Generalised Lomb--Scargle  periodogram of the bisector velocity span of the SOPHIE observations. A clear peak is seen at P=4.58 days, in agreement with the rotational period determination by \citet{mazeh2015}, whose 1-$\sigma$ range is shown in grey.}
\label{fig.bis}
\end{figure}

The observed CCFs are broad; they have a full width at half maximum of $19.9\pm0.2$~\kms, which corresponds to a projected rotational velocity $v\sin i_\star = 12.3 \pm 1.0$~\kms\ \citep{boisse2010}. From two high-resolution HIRES spectra \citet{dawson2014} measured  $v\sin i_\star = 14.4 \pm 1.3$~\kms, in agreement with the SOPHIE estimate. Using these estimates and the amplitude of the photometric variability, the expected stellar jitter produced by stellar spots on this rapidly rotating F7 star can be estimated using a simplified formula for the amplitude of the Rossiter--McLaughlin effect\footnote{$\Delta RV [\ms]\approx 1.1\; \vsini[\kms]\; \Delta F [mmag]$}. The \Kepler\ light curve presents a variability of $\sim2$~mmag peak-to-peak amplitude, which leads to an estimate radial velocity jitter of $\sim27$~\ms. Additionally, the time series of the bisector velocity span shows a clear periodicity at 4.58 days (Fig.~\ref{fig.bis}), in agreement with the measured rotational period from Kepler photometry \citep{mazeh2015}.

For our analyses we also used the 20 HIRES \citep{vogt1994} radial velocities presented by \citet{dawson2014}, acquired between April and September 2012, with one additional point on August 2013. Their typical internal precision is of the order of $\pm12$~\ms. However, \citet{dawson2014} report a scatter of 40~\ms\ around their best-fit model, which they attribute to stellar effects. By comparison with the HIRES data set, the new SOPHIE data provide a longer time span with an improved time sampling.

\begin{figure}[t]
    \begin{subfigure}[t]{0.47\textwidth}  
        \caption{HIRES+SOPHIE}
        \vspace{-0.25cm}
        \includegraphics[width=0.95\textwidth]{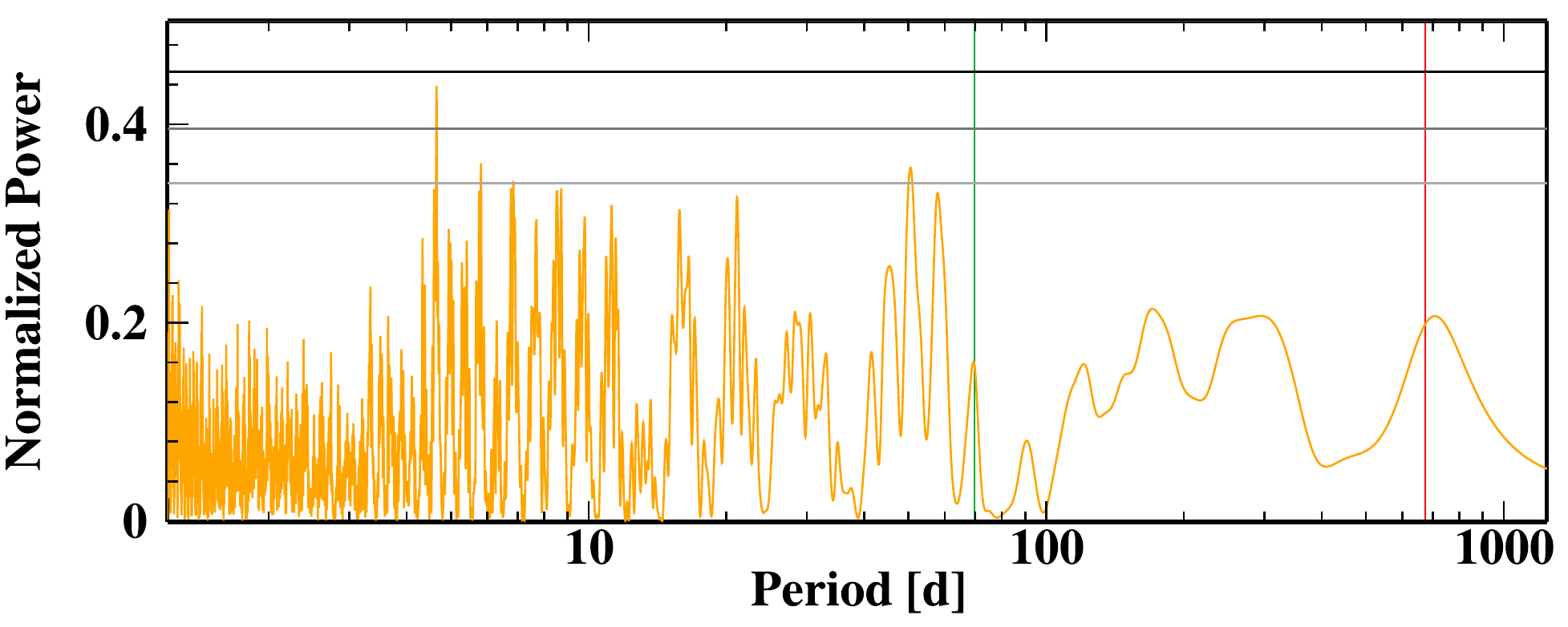}
        \vspace{0.01cm}
    \end{subfigure}
    
    \begin{subfigure}[t]{0.47\textwidth}  
        \caption{HIRES}
        \vspace{-0.25cm}
        \includegraphics[width=0.95\textwidth]{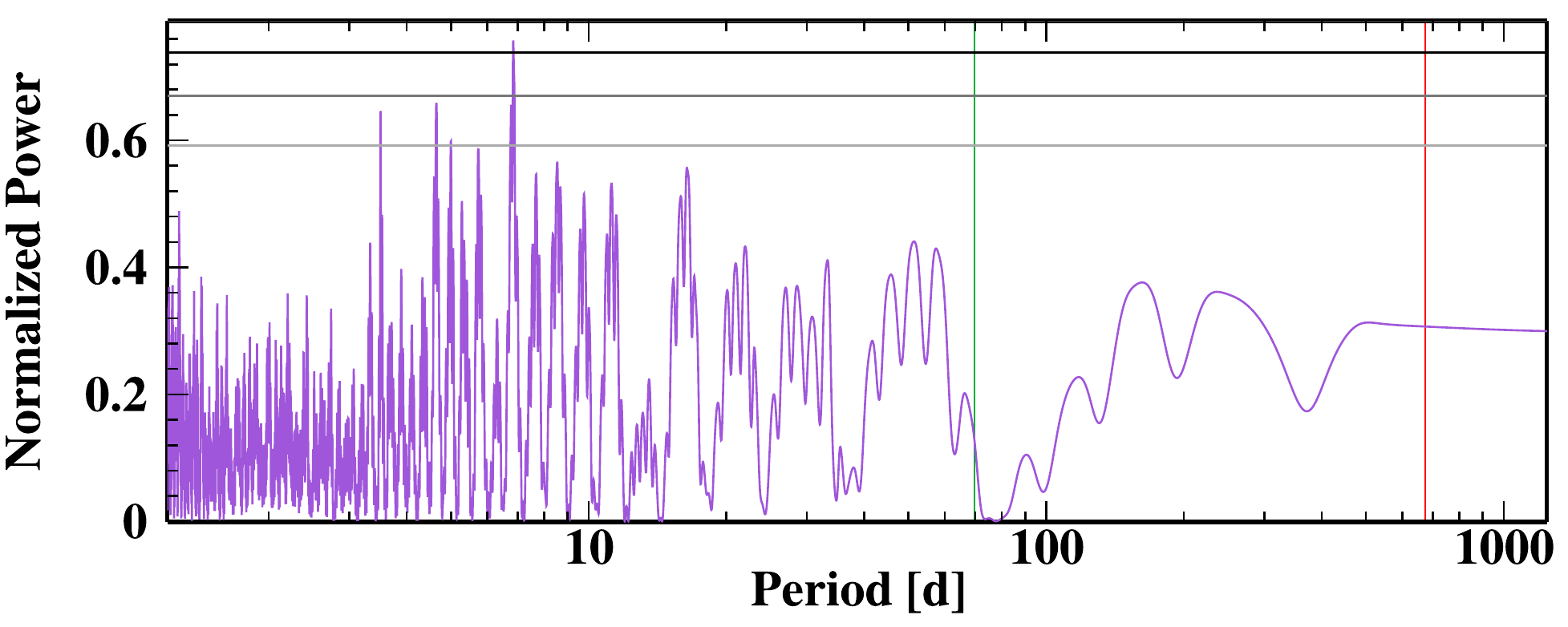}
        \vspace{0.01cm}
    \end{subfigure}
    
    \begin{subfigure}[t]{0.47\textwidth}  
        \caption{SOPHIE}
        \vspace{-0.25cm}
        \includegraphics[width=0.95\textwidth]{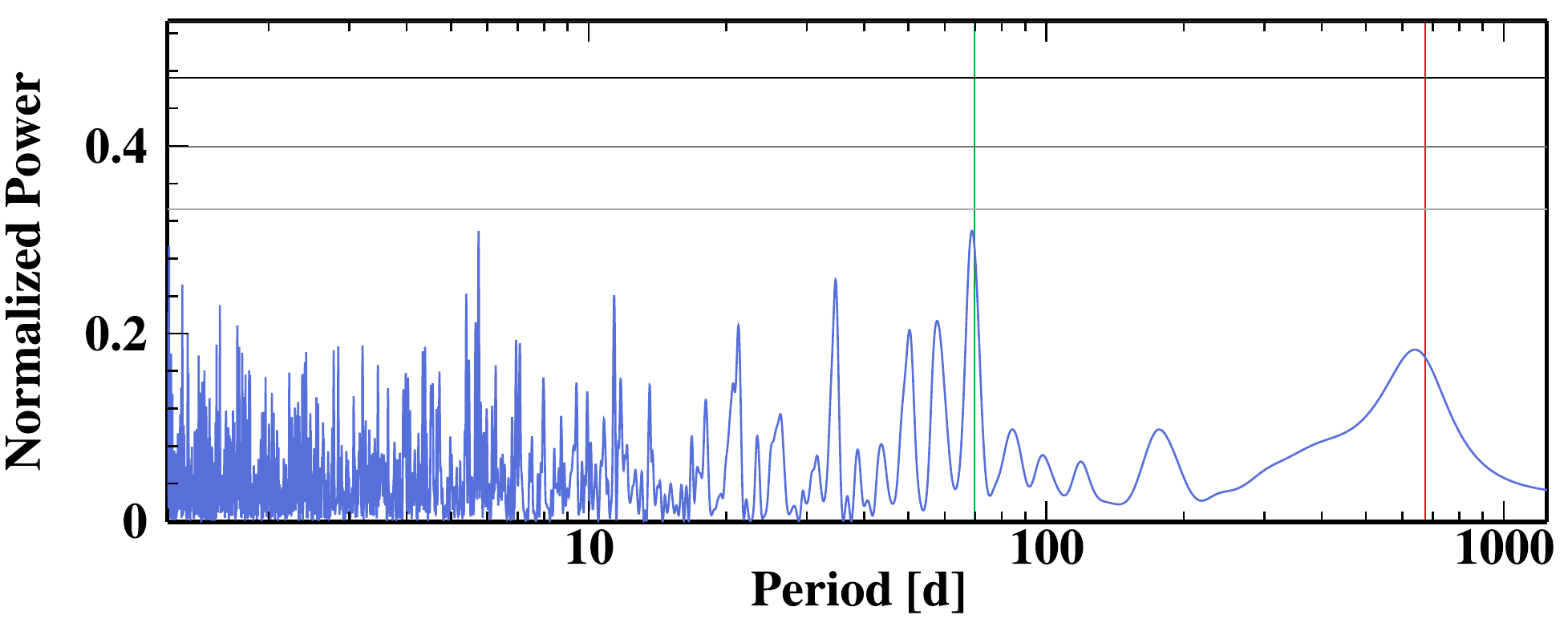}
        \vspace{0.01cm}
    \end{subfigure}
    
    \begin{subfigure}[t]{0.47\textwidth}
        \caption{SOPHIE -- Keplerian planet~b}
        \vspace{-0.25cm}
        \includegraphics[width=0.95\textwidth]{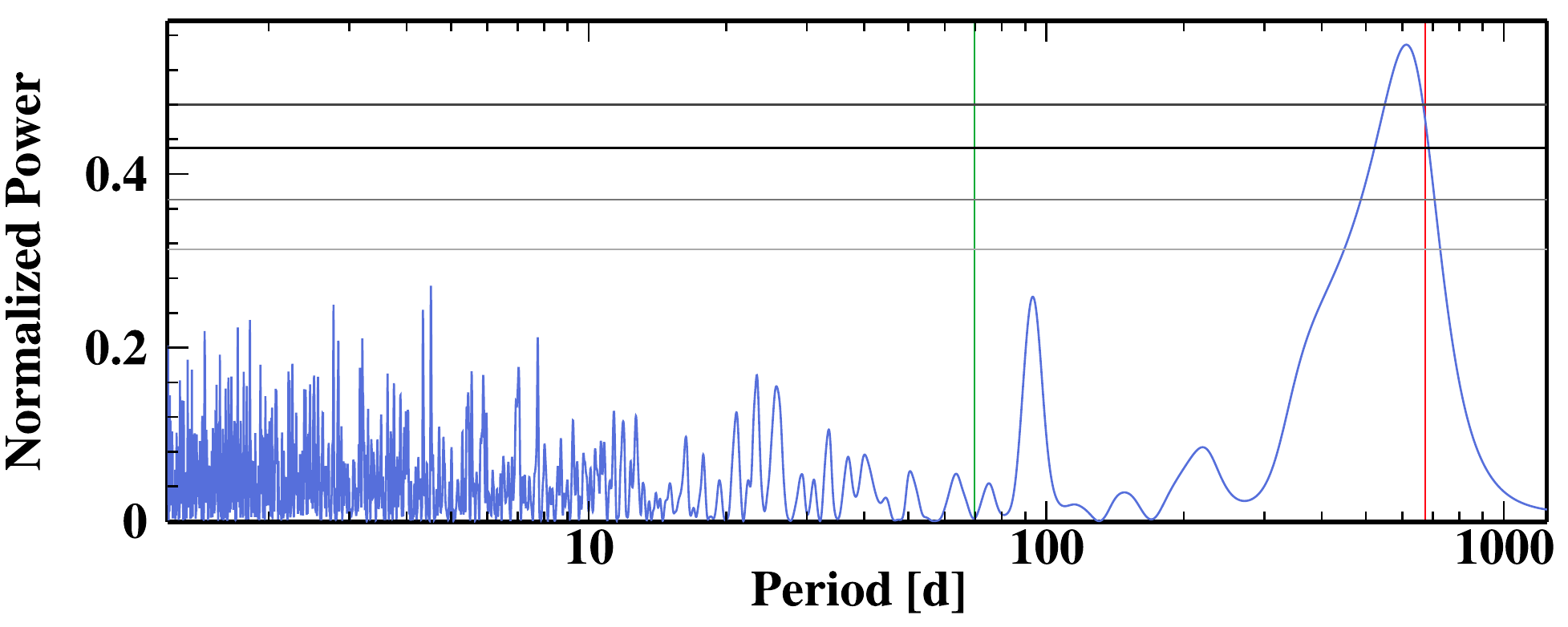}
        \vspace{0.01cm}
    \end{subfigure}
    
    \begin{subfigure}[t]{0.47\textwidth}
        \caption{SOPHIE -- Keplerian planets~b and c}
        \vspace{-0.25cm}
        \includegraphics[width=0.95\textwidth]{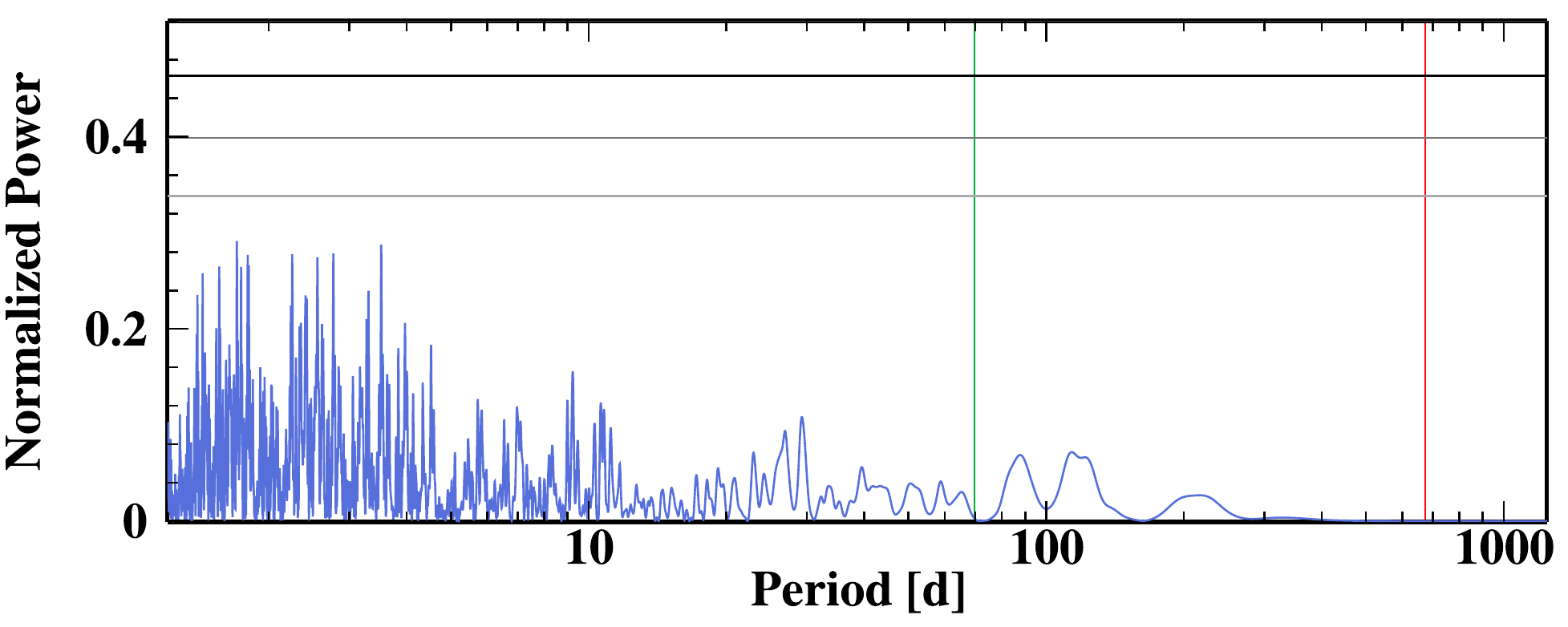}
        \vspace{0.01cm}
    \end{subfigure}
    
    \caption{Generalised Lomb--Scargle  periodograms of the radial velocities. The green and red vertical lines mark the period of planet~b and c, respectively. The grey horizontal lines represent the 50\%, 10\%, and 1\% p-value levels. From top to bottom: HIRES and SOPHIE data; HIRES velocities; SOPHIE velocities; SOPHIE after substraction of Keplerian curve at the period of planet~b; SOPHIE velocities after substraction of a two-Keplerian model with the periods of planets~b and c. The periodogram plots are provided by DACE (\protect\url{dace.unige.ch}).}
    \label{fig.LSRV}
\end{figure}

\section{Analysis}

\subsection{First analysis: Keplerian model}\label{sect.keplerian}

We first fitted the radial velocities with a two-planet Keplerian model, i.e. neglecting the mutual gravitational interactions between both planets. 
The goal here was to have a first idea of the information included in the radial velocity data set. 

Figure~\ref{fig.LSRV} shows Generalised Lomb--Scargle (GLS) periodograms \citep{zechmeister2009} of the radial velocities. In each panel the periods of the two planets are indicated with the green and the red lines (70 and 673~days, respectively). The first panel shows the periodogram of the mean-corrected SOPHIE and HIRES velocities together. It reveals no significant signals, but there is a peak close to the rotational period of the star, 4.5 days. The periodogram of the HIRES data is presented in second panel of Fig.~\ref{fig.LSRV}. No significant power is seen at the periods of any of the planets, and there is a forest of peaks around the rotational period, including a seemingly significant one around 7 days. We cannot explain this signal easily, but as it is only detected in the HIRES data, we are inclined to ascribe it to instrument systematics (see also discussion at the end of the section). The third panel of Fig.~\ref{fig.LSRV} presents the periodogram of SOPHIE data only. The $p$-values of the peaks at the planet periods, computed by randomly shuffling the model residuals, remain above the customary 1\%. However, when the Keplerian orbit is eccentric, the GLS periodogram power is transferred partially to the harmonics of the orbital period, leading to a reduced peak amplitude. More robust methods should be used to assess the question of the significance of these signals in detail, but this is beyond the scope of this paper. 

The residuals of a Keplerian fit including Kepler-419b only show a strong signal at the period of Kepler-419c, with a $p$-value smaller than 1\%  (Fig.~\ref{fig.LSRV}, fourth panel). We conclude that the outer, non-transiting planet Kepler-419c, which was only predicted from TTVs, is detected in the SOPHIE radial velocities. In the same way, if the effect of the outer planet is removed from the SOPHIE data, a significant peak ($p$-value slightly higher than 1\%)  appears at the period of the inner transiting planet.

\begin{table}[h]
\caption{Two-Keplerian fit to the SOPHIE data (orbital period, time of conjunction, eccentricity, argument of pericentre, radial velocity semi-amplitude, minimum mass, and systemic velocity).}
\begin{center}
\begin{tabular}{ccccc}
\hline
\hline
Keplerian&&Kepler-419\,b&Kepler-419\,c\\
\hline
P&[d]&69.74$\pm$0.10&667$\pm$23\\
T$_{\rm c}$&[BJD]&2\;454\;959.2$\pm$2.8&2\;455\;491$\pm$41\\
e&&0.800$\pm$0.036&0.130$\pm$0.097\\
$\omega$&[\degree]&86$\pm$12&283$\pm$38\\
K&[\ms]&181$\pm$20&147$\pm$18\\
$M_{\rm p}\sin{i}\;^\dagger$&[\Mjup]&2.71$\pm$0.40&7.8$\pm$1.0\\
\hline
$\gamma_{\rm SOPHIE}$&[\kms]&\multicolumn{2}{c}{-25.877$\pm$0.011}\\
\hline
\end{tabular}
\end{center}
\begin{list}{}{}
\item {\bf{Notes.}}
  $^{(\dagger)}$ Using as stellar mass $M_\star=1.40_{-0.08}^{+0.06}~\Msun$ from \citet{dawson2014}. 
\end{list}
\label{default}
\end{table}

Table~1 shows the results of the two-planet Keplerian model fit \citep[using DACE\footnote{\url{dace.unige.ch}};][]{delisle2016} to the SOPHIE data alone. The parameters of the Keplerian orbits are both in agreement with the parameters reported by \citet{dawson2014} and with our photodynamical model presented in Sect.~\ref{final}; in particular, the orbital periods and the phases agree.

The residuals of the two-Keplerian fit show no additional signals (Fig.~\ref{fig.LSRV}, lower panel). The scatter of the residuals is  38~\ms, in good agreement with the precision of the observations and the expected stellar jitter\footnote{$\sqrt{\left(28~\ms\right)^2 + \left(27~\ms\right)^2} \simeq 39~\ms$}. This is similar to the dispersion measured in the HIRES residuals after a similar fit. As the HIRES data have a smaller internal dispersion (12~\ms), this hints at some level of systematics in the HIRES data.

\subsection{Photodynamical modelling without radial velocities}\label{onlyPH}

\begin{figure}
\includegraphics[width=0.48\textwidth]{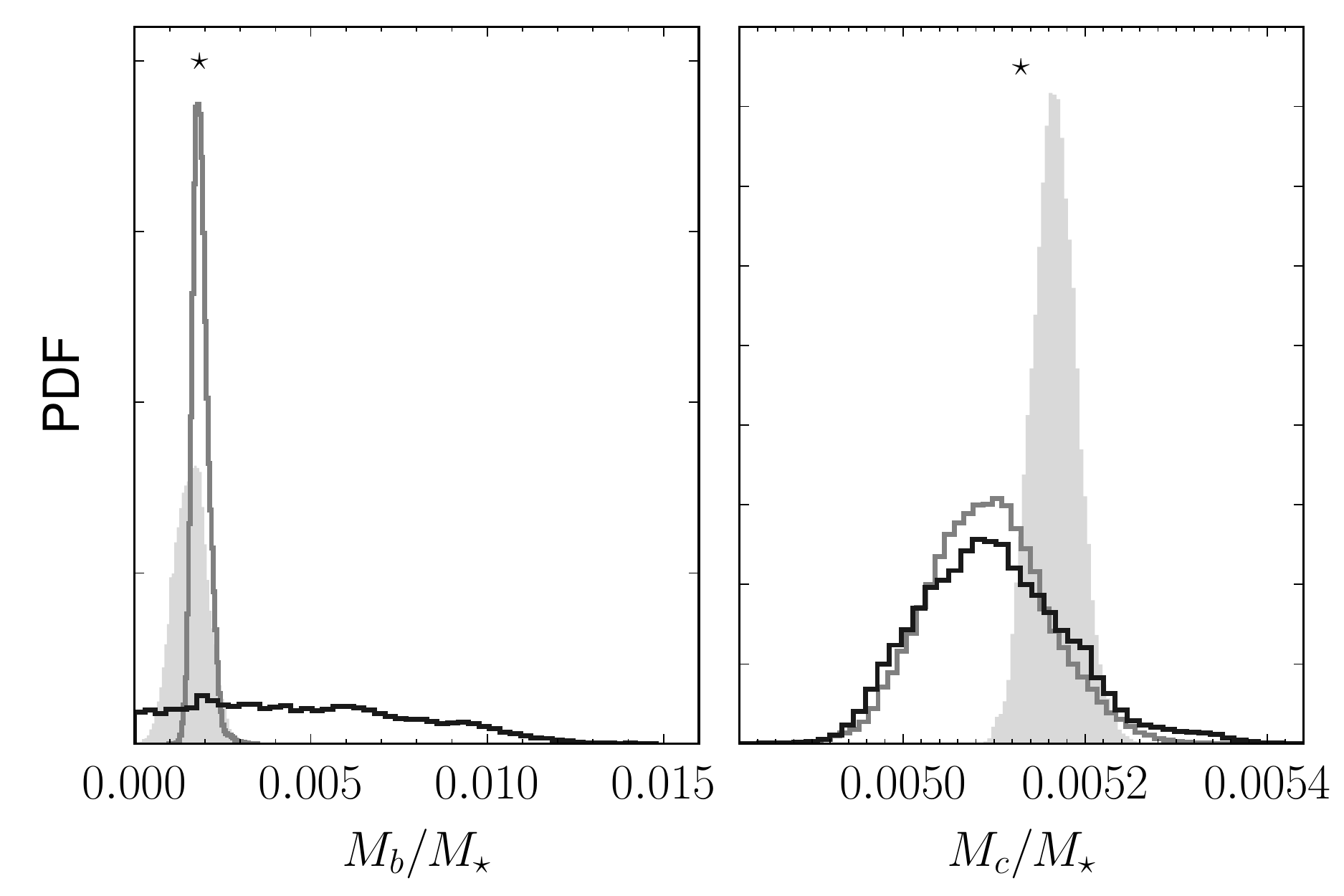}
\caption{Posteriors of planet-to-star mass ratios for planet~b (left) and planet~c (right). The black histograms correspond to the analysis of the photometric data alone, while the grey histogram includes radial velocities. The filled grey histograms show the PDF for the masses of planets~b and c respectively for the analysis of the simulated light curve with 54 transits and no radial velocities.}
\label{fig.mr}
\end{figure}

The photodynamical analysis of the \Kepler\ light curve, including the interaction between the planets, allows  the planet-to-star mass ratios to be constrained. The details on the modelling are given in Section~\ref{final}. Here we neglected effects related to the time of travel of light, which we confirmed does not change the result significantly (see Section~\ref{final}). As a consequence, the model used in this section does not depend on the sizes and masses of the star and planets, only on their densities. In Fig.~\ref{fig.mr} we present the posterior distributions of the mass ratios of planets~b and c relative to the star. Contrary to the commonly accepted notion that TTVs of a given planet are completely insensitive to its own mass, the distribution in Fig.~\ref{fig.mr} (left panel, black histogram) shows that an upper limit for the mass ratio of planet~b can be set (0.010 at 95\% confidence level). We study the constraints on the mass of planet~b in detail in Sect.~\ref{sect.massconstrains}. On the other hand, the planet-to-star mass ratio of planet~c is well constrained using \Kepler\ photometry data alone (Fig.~\ref{fig.mr}, right panel).

\subsection{Final analysis: photodynamical model}\label{final}

Finally, we employed a photodynamical model of the observed photometry and radial velocity measurements, accounting for the gravitational interactions of all the known components of the system. Our model is described in detail in \citet{almenara2015, almenara2016}. In brief, we obtain the positions and velocities of the three bodies in the system in time through numerical integration of the system. The sky-projected positions are used to compute the light curve \citep{mandelagol2002} using a quadratic limb-darkening law \citep{manduca1977}. To account for the integration time, the model was oversampled by a factor of 30 and 3, for the long- and short-cadence data respectively, and then binned back to match the cadence of the data points \citep{kipping2010}. The line-of-sight projected velocity of the star issued from the integration is used to model the radial velocity measurements (i.e. we do not assume Keplerian motion).

\begin{figure*}
\hspace{-0.2cm}\includegraphics[height=5.7cm]{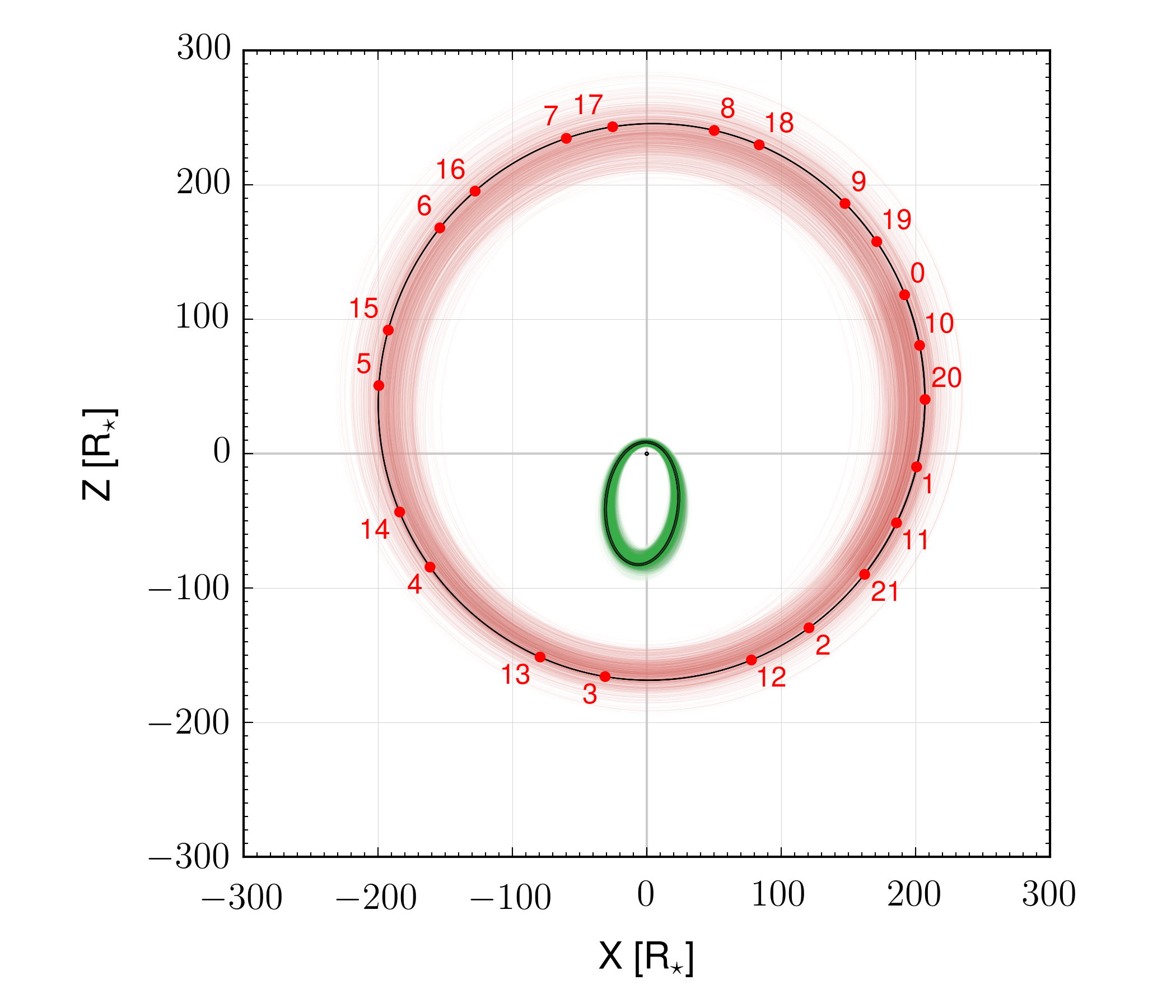}\hspace{-0.5cm}\includegraphics[height=5.7cm]{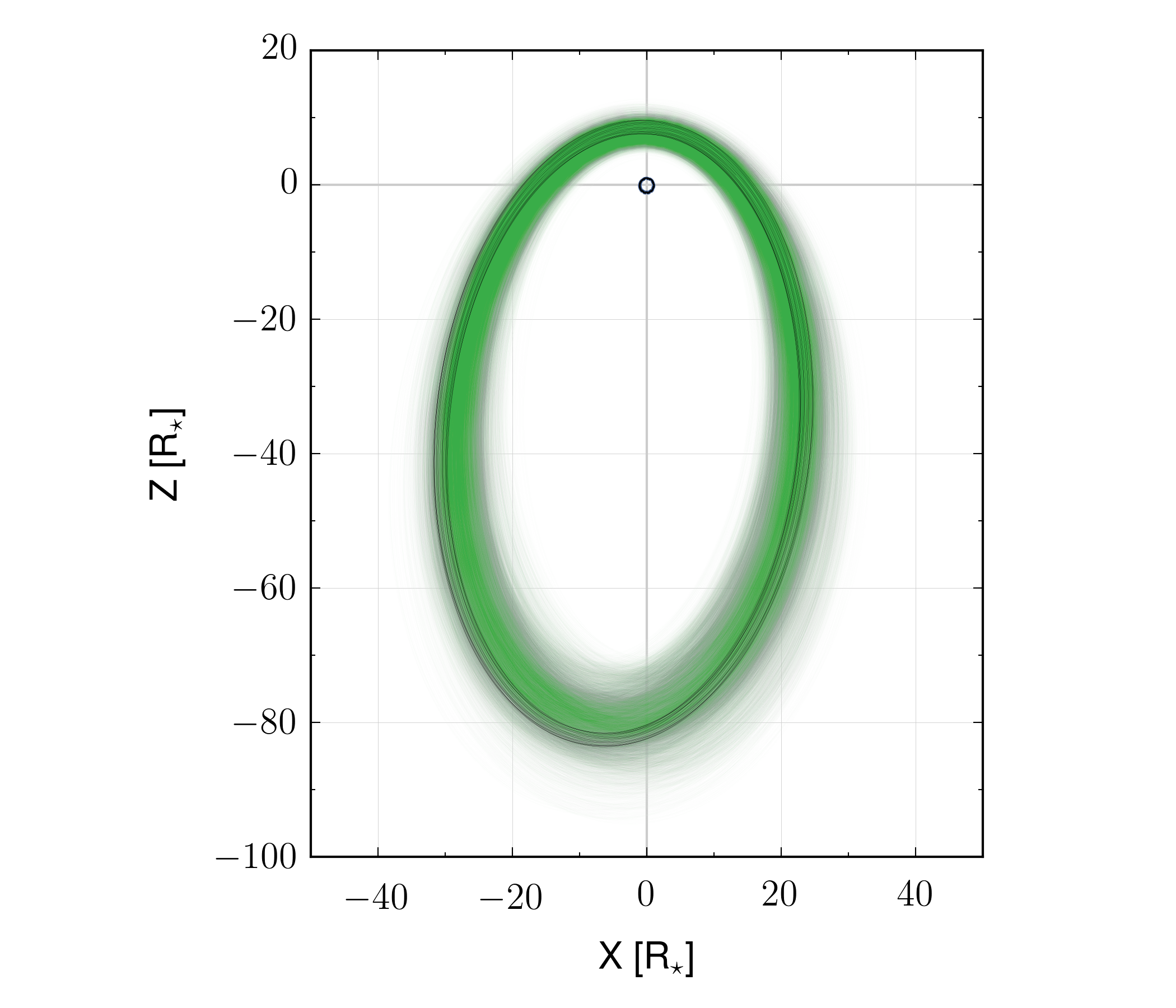}\hspace{-0.5cm}\includegraphics[height=5.7cm]{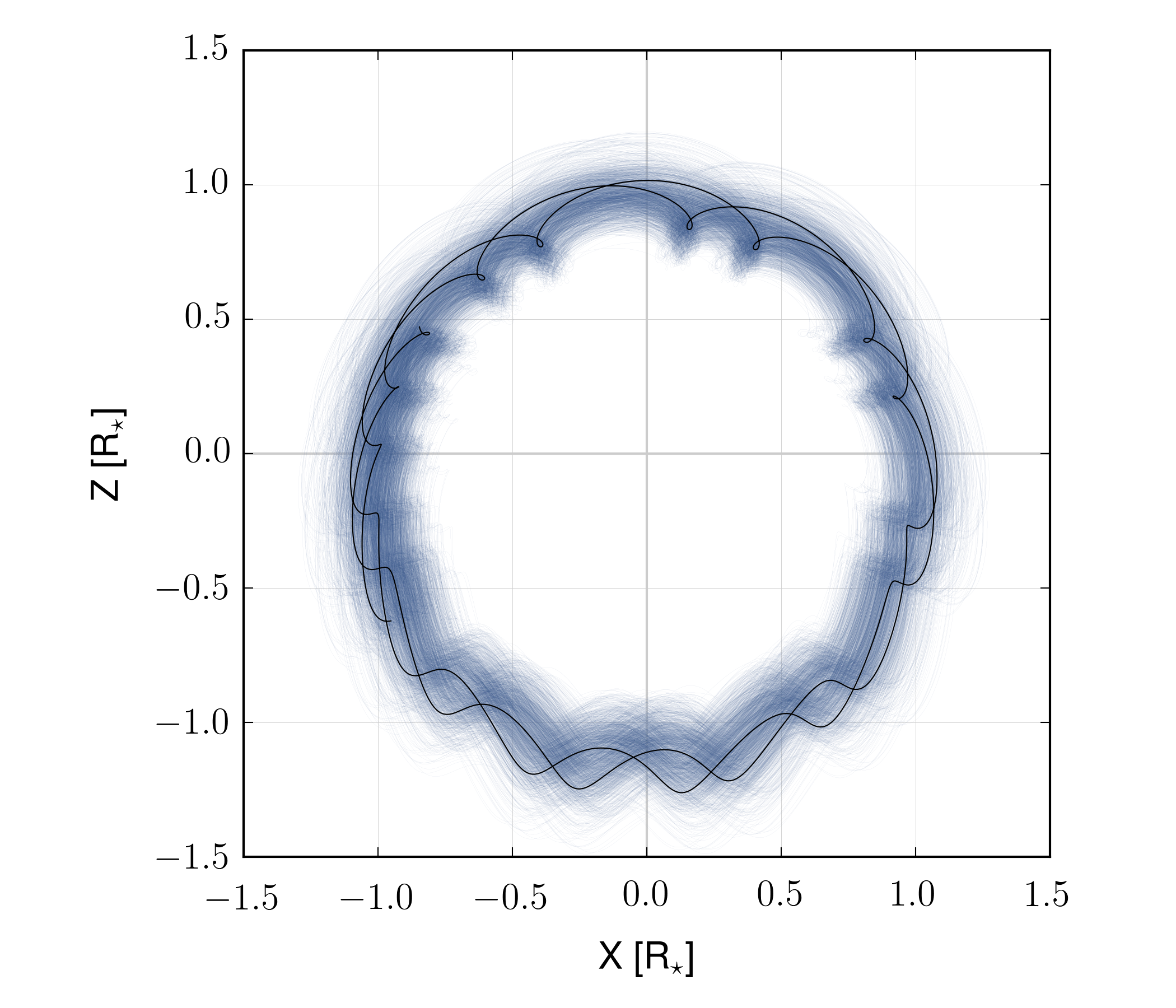}\\
\vspace{-0.5cm}
\caption{Orbital projections during the \Kepler\ observations for planet~b (green), planet~c (red), and the star (blue). The origin is the system barycentre, the movement is clockwise, and the orbits are projected in the X-Z reference plane, with the positive Z-axis pointing towards the observer. The two rightmost panels are successive zooms of the panel on the left. A thousand random orbits are drawn form the posterior samples, and the MAP is shown as a black orbit. The red points in the leftmost panel mark the position of planet~c on the MAP orbit at the epochs of transits of planet~b, numbered accordingly.}
\label{fig.orbits}
\end{figure*}

We used the n-body code \reb \citep{rein2012} with the \whf integrator \citep{rein2015} and an integration step of 0.01~days, which results in a maximum error of 4~\cms\ and 1~ppm for the radial velocity and photometric model, respectively \citep[which also takes into account the oversampling factor,][]{kipping2010}. We included the light-time effect \citep{irwin1952}, which has an amplitude of $\sim$5~s on the TTVs, corresponding to a displacement of the star by around 2 stellar radii (see Fig.~\ref{fig.orbits}). This is small compared to the timing precision of individual transits, and we have checked that the results are not significantly different when the effect is not included, as in Sect.~\ref{onlyPH}. The model is parametrised using osculating astrocentric asteroidal orbital elements (Table~\ref{table.results}) at the time immediately before the first transit observed by \Kepler\, $t_{\mathrm{ref}} = 2\;454\;958$~BJD$_{\mathrm{TDB}}$, given in Barycentric Dynamical Time. Due to the symmetry of the problem, we fixed the longitude of the ascending node of the interior planet $\Omega_{\mathrm{b}}$ at $t_{\mathrm{ref}}$, we limited the inclination of the outer one $i_c<90$\degree, and rejected models where planet~c transits\footnote{The transits of planet~c were ruled out by \citet{dawson2014}.}. 

Our model has 26 free parameters. In addition to the physical parameters, we considered a radial velocity offset for each instrument with respect to the systemic velocity (assumed to be zero), a global light curve normalisation factor for long- and short-cadence data, and a multiplicative jitter parameter for each data set. A non-informative uniform prior distribution was chosen, and the joint posterior distribution was sampled using the \emcee\ algorithm \citep{goodmanweare2010, emcee}. To minimise correlations, the combinations of parameters listed in Table~\ref{table.results} were used for the Markov chain Monte Carlo (MCMC) algorithm. We ran \emcee\ with 100~walkers from a starting point based on the results of \citet{dawson2014}. We ran 2.4$\e{6}$ steps of the \emcee\ algorithm, and used the last 100\;000 steps for the final inference.

\section{Results}

\begin{figure}
\includegraphics[width=0.50\textwidth]{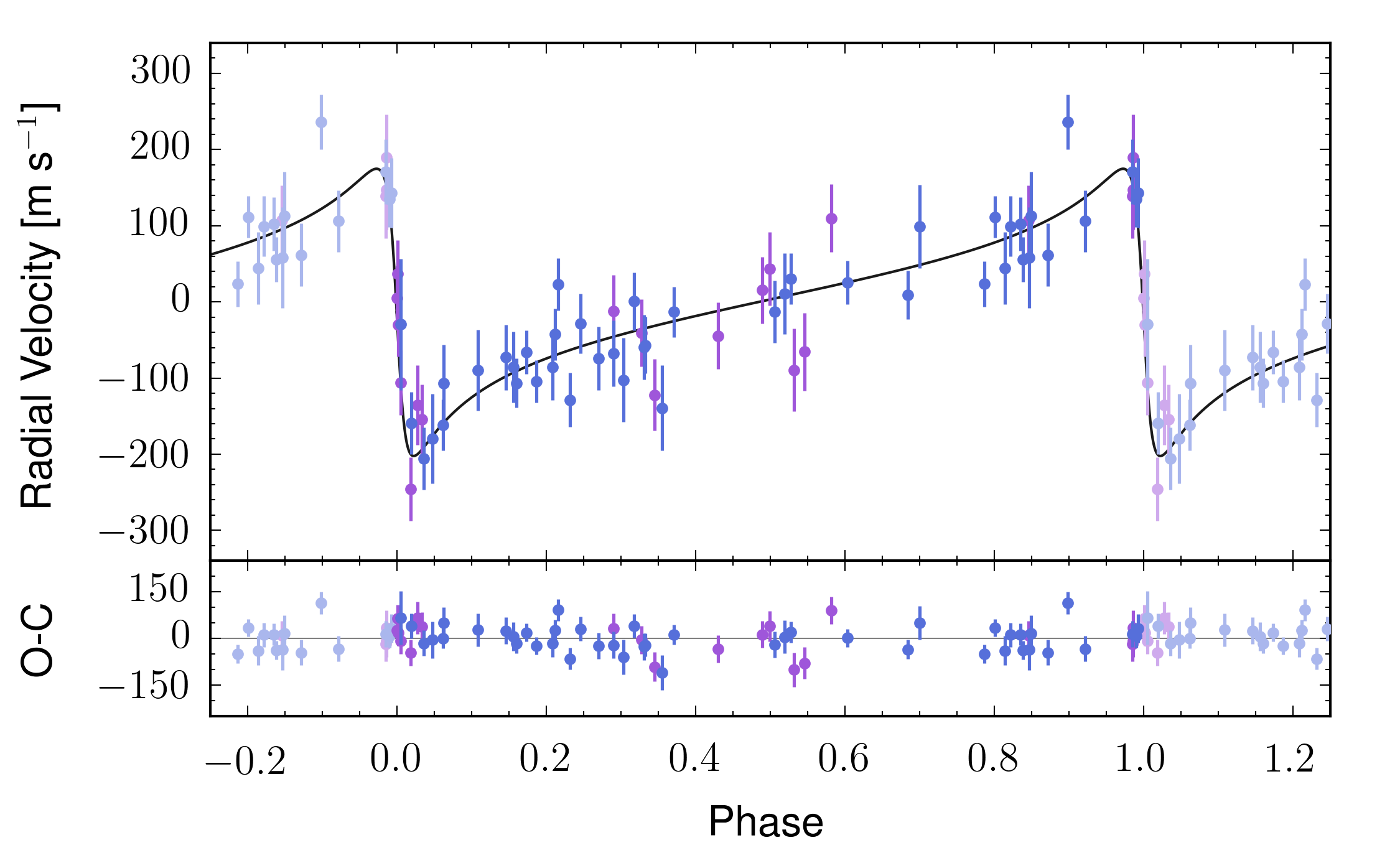}
\includegraphics[width=0.50\textwidth]{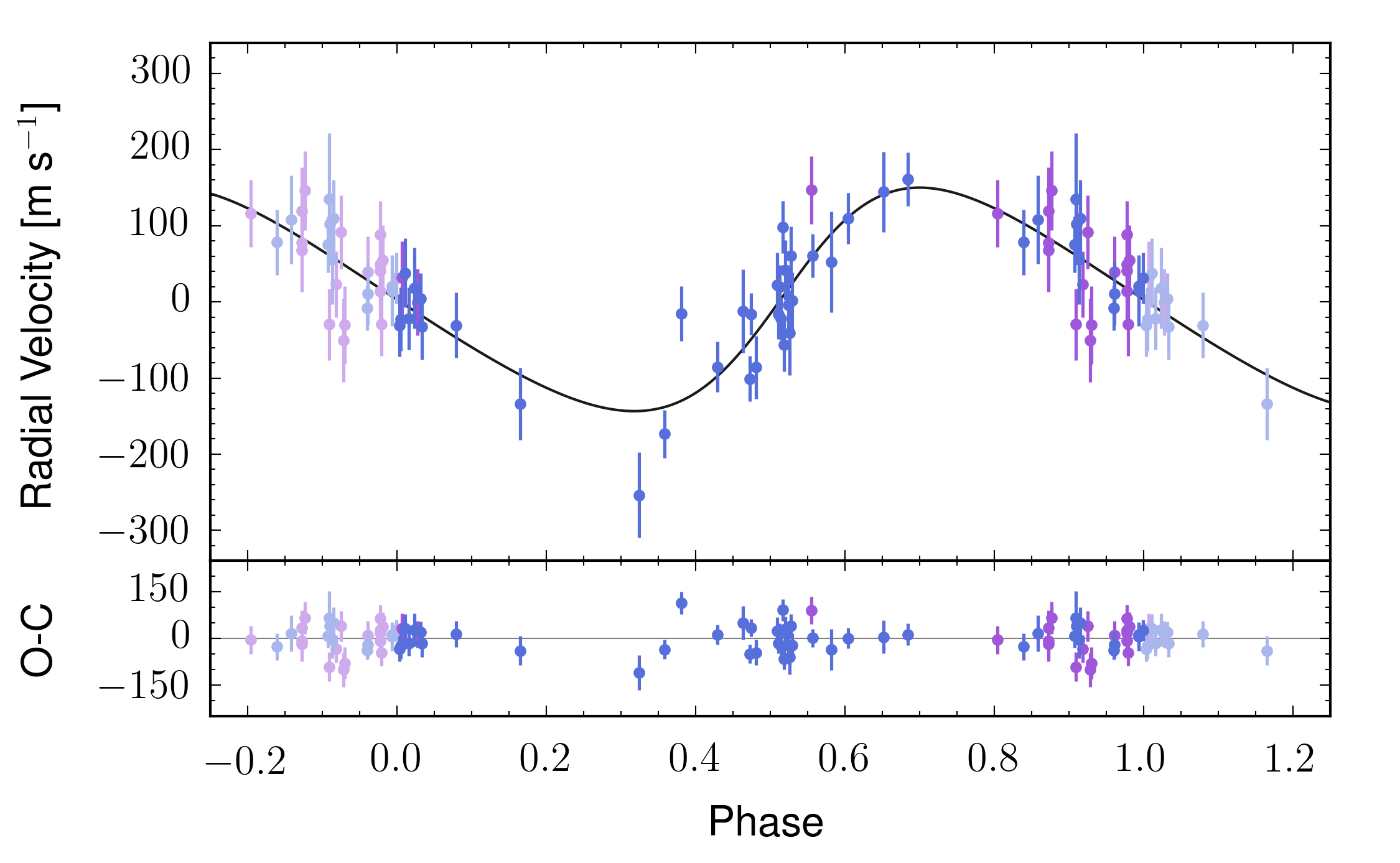}
\caption{Phase-folded radial velocities for planet~b (top) and c (bottom). The points are colour-coded as in Fig.~\ref{fig.RV} (blue for SOPHIE and purple for HIRES). In the photodynamical model the contribution to RV signal coming from individual planets cannot be separated. Therefore, these figures are constructed by fitting a two-Keplerian model to the photodynamical MAP model radial velocities, and subtracting the individual contribution of each planet from the data. The uncertainties include the MAP jitter value.}
\label{fig.RVphase}
\end{figure}

In Table~\ref{table.results} we list the maximum a posteriori (MAP) estimate, the median, the 68\% credible interval (CI), and the 95\% highest density interval (HDI) of the inferred system parameters marginal distributions\footnote{The one- and two-dimensional projections of the posterior sample are shown in Fig.~\ref{fig.pyramid}}. The MAP model and credible regions are plotted in Figures~\ref{fig.PH} and \ref{fig.RV}. Figure \ref{fig.RVphase} presents the RV measurements, phase-folded to the best-fit Keplerian period to the MAP photodynamical model.

In Fig.~\ref{fig.mr} we present the histogram of the marginal distribution of the planet-to-star mass ratios. The distribution for planet~c is only slightly affected by the inclusion of the radial velocities. On the other hand, the distribution for planet~b is drastically improved by the inclusion of the SOPHIE RVs. By self-consistent modelling of photometry and radial velocity, absolute masses and radii are inferred without resorting to theoretical stellar model. The radii are measured to a precision of 12\% and 13\% for the star and planet~b, respectively. The masses were determined to a precision of 35\%, 24\%, and 35\% for the star, planet~b, and planet~c, respectively. The precision on the densities, also derived dynamically, are 19\% and 26\% for the star and planet~b, respectively.

\begin{figure}
\hspace{-0.4cm}\includegraphics[width=0.52\textwidth]{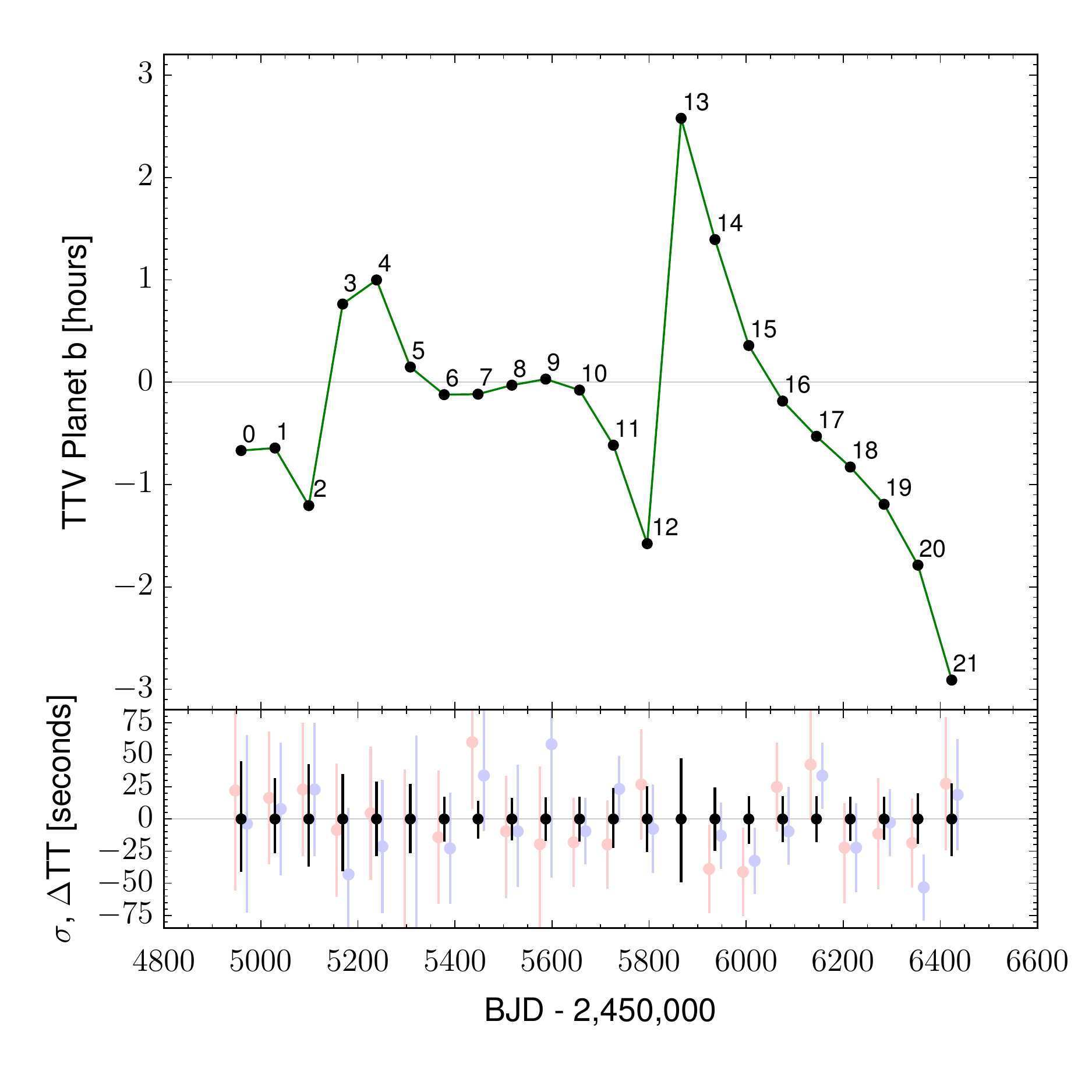}
\vspace{-0.7cm}
\caption{Posterior TTVs of Kepler-419b, computed relative to a linear ephemeris computed using the median values of each epoch during \Kepler\ observations (BJD$_{\rm TDB}$ = 2\;454\;959.358(23) + 69.728\;8(21)\;$\times$\;Epoch, where the errors are indicated in parentheses). A thousand random draws from the posterior distribution are used to estimate the TTV median value and its uncertainty. In the upper panel the median TTV values are labelled with the corresponding epoch number (0 is the first transit observed by \Kepler). In the lower panel, the posterior median transit timing value is subtracted to visualise the uncertainty of the distribution. The two sets of transit times derived in \citet{dawson2014} are shown in light red (TAP) and light blue (GP), slightly offset in the x-direction for clarity. The median transit time was subtracted from each epoch to allow for comparison with our results.}
\label{fig.TTV}
\end{figure}

Figure~\ref{fig.TTV} shows the posterior of the TTV, obtained as the mean of the first and fourth contacts interpolated from the sky-projected planet–star separation. The uncertainty in the transit times is not constant. \citet{almenara2015} showed that the TTV posterior distributions are typically narrower around the mid-time of the observations. In addition to this effect, in this case we have a combination of long- and short-cadence transits, giving larger uncertainties for long-cadence epochs 0 to 9 and epoch 21. The uncertainty is reduced for the short-cadence transits (epochs 10 to 20), but there is  an increase due to a missed transit at epoch 13. The mean uncertainty of the transit times derived with the photodynamical modelling is two times smaller than any of the timing sets computed on individual transits by \citet{dawson2014}. Clearly, our values rely on the three-body system hypothesis. However, this seems to be a pertinent assumption, as our derived transit times are in agreement with the values presented in \citet{dawson2014}. If more bodies are present in the system, they seem to be dynamically irrelevant at these timescales.

The data favour a model where the planets have low true mutual inclination ($i_\mathrm{rel}$)\footnote{$\cos i_{rel}= \cos i_b\cos i_c + \sin i_b\sin i_c \cos\left({\Omega_b-\Omega_c}\right)$}. We obtained $i_\mathrm{rel} = 6.6^{+6.5}_{-4.0}$~\degree\ (median and 68.3\% CI) at $t_{\mathrm{ref}}$, with the mode at 2.5\degree, and an upper 99\% confidence limit of 23.5\degree, in agreement with \citet{dawson2014}. These authors reported that the mutual inclination was below 21\degree, at a 91\% confidence level. 

\begin{table*}
  \tiny
\renewcommand{\arraystretch}{1.1}
\centering
\cprotect\caption{Inferred system parameters: MAP, 95\% HDI, posterior median and 68.3\% CI for the model-free photodynamical analysis. The last column lists the median and 68.3\% CI for the parameters whose precision is improved by using theoretical stellar models. The astrocentric orbital elements are given for the reference time $t_{\mathrm{ref}} = 2\;454\;958$~BJD$_{\mathrm{TDB}}$. We use the nominal units established by the recent 2015 IAU B3 resolution on Recommended Nominal Conversion Constants for Selected Solar and Planetary Properties \citep{iau2015}, which are listed in the table notes.} \label{table.results}
\begin{tabular}{lccccc}
\hline
Parameter & & MAP &  95\% HDI & Median  &Stellar models \\
&  & & & and 68.3\% CI & median and 68.3\% CI \\
\hline
\rule{0pt}{2ex}
\emph{\bf Star} \smallskip\\

Stellar mass, $M_\star$                            & [\Msun]                 & 1.581       & [0.612, 2.347]             & 1.39$\pm$0.48          & 1.438$\pm$0.053 \\
Stellar radius, $R_\star$$^{\bullet}$              & [\Rnom]                 & 1.812       & [1.387, 2.237]             & 1.80$\pm$0.22          & 1.81$\pm$0.12 \\
Stellar mean density, $\rho_{\star}$$^{\bullet}$   & [$\mathrm{g\;cm^{-3}}$] & 0.3745      & [0.2321, 0.4539]           & 0.335$\pm$0.062        & \\
Surface gravity, \logg\                            & [cgs]                   & 4.1205      & [3.9231, 4.1934]           & 4.072$\pm$0.075        & 4.080$\pm$0.046 \smallskip\\

$q_1$$^{\dagger,\bullet}$                          &                         & 0.2519      & [0.1464, 0.3647]           & 0.245$\pm$0.059        & \\
$q_2$$^{\dagger,\bullet}$                          &                         & 0.369       & [0.189, 0.645]             & 0.40$\pm$0.13          & \\
Linear limb darkening, $u_{\mathrm{a}}$            &                         & 0.3704      & [0.2563, 0.5429]           & 0.392$\pm$0.074        & \\
Quadratic darkening, $u_{\mathrm{b}}$              &                         & 0.131       & [-0.136, 0.351]            & 0.10$\pm$0.12          & \medskip\\

\emph{\bf Planet~b} \smallskip\\

Semi-major axis, $a_b$                               & [au]                    & 0.3865      & [0.2931, 0.4489]           & 0.371$\pm$0.040        & 0.3745$\pm$0.0046 \\
Eccentricity, $e_b$                                  &                         & 0.8070      & [0.7857, 0.8462]           & 0.817$\pm$0.016        & \\
Inclination, $i_b^{\bullet}$               & [\degree]               & 87.372      & [85.560, 88.482]           & 87.04$\pm$0.72         & \\
Argument of pericentre, $\omega_b$                   & [\degree]               & 95.23       & [89.65, 97.91]             & 94.0$\pm$2.2           & \\
Longitude of the ascending node, $\Omega_b$          & [\degree]               & 180$^{\ast}$ &  &  & \\
Mean anomaly, ${M_0}_b$                                & [\degree]               & 352.800     & [352.668, 353.158]         & 352.90$\pm$0.12        & \smallskip\\

Radius ratio, $R_b/R_\star^{\bullet}$ &                         & 0.063551    & [0.062505, 0.064722]       & 0.06359$\pm$0.00056    & \\
Mass ratio, $M_b/M_\star^{\bullet}$   &                         & 0.001852    & [0.001459, 0.002360]       & 0.00186$\pm$0.00025    & 0.00183$\pm$0.00012 \\
Scaled semi-major axis, $a_b/R_{\star}$              &                         & 45.86       & [39.16, 49.05]             & 44.2$\pm$2.6           & \\
${T_0}_b'^{\bullet}$\;-\;2\;450\;000                 & [BJD$_{\mathrm{TDB}}$]  & 4959.331882 & [4959.330706, 4959.332819] & 4959.33177$\pm$0.00054 & \\
$P'_b$ $^{\bullet}$                                   & [d]                     & 69.79631    & [69.78350, 69.81399]       & 69.7968$\pm$0.0087     & 69.7960$\pm$0.0042 \\
$K'_b$                                               & [\ms]                   & 188.8       & [162.7, 209.1]             & 186$\pm$12             &  \smallskip\\
$\sqrt{e_b}\cos{\omega_b}^{\bullet}$                 &                         & -0.0818     & [-0.1234, 0.0072]          & -0.063$\pm$0.035       & \\
$\sqrt{e_b}\sin{\omega_b}^{\bullet}$                 &                         & 0.89461     & [0.88458, 0.91749]         & 0.9007$\pm$0.0086      & \\

Planet mass, $M_b$                      &[\Mjup]                  & 3.068       & [1.584, 4.002]             & 2.71$\pm$0.66          & 2.77$\pm$0.19 \\
Planet radius, $R_b$                    &[\RJnom]                 & 1.121       & [0.857, 1.392]             & 1.11$\pm$0.14          & 1.120$\pm$0.084 \\
Planet mean density, $\rho_b$           &[$\mathrm{g\;cm^{-3}}$]  & 2.702       & [1.443, 3.688]             & 2.43$\pm$0.62          & \\
Planet surface gravity, $\log$\,$g_b$   &[cgs]                    & 3.7820      & [3.5999, 3.8628]           & 3.731$\pm$0.066        & \medskip\\

\emph{\bf Planet~c} \smallskip\\

Semi-major axis, $a_c$                               & [au]                    & 1.752       & [1.328, 2.037]             & 1.68$\pm$0.18          & 1.697$\pm$0.020 \\
Eccentricity, $e_c$                                  &                         & 0.17973     & [0.17631, 0.18257]         & 0.1793$\pm$0.0017      & \\
Inclination, $i_c$ $^{\bullet}$                       & [\degree]               & 85.72       & [83.64, 89.74]             & 87.0$\pm$2.0           & \\
Argument of pericentre, $\omega_c$                   & [\degree]               & 276.75      & [272.19, 279.03]           & 275.7$\pm$1.8          & \\
Longitude of the ascending node, $\Omega_c$ $^{\bullet}$ & [\degree]            & 184.77      & [170.98, 201.27]           & 185.4$\pm$7.6          & \\
Mean anomaly, ${M_0}_c$                                & [\degree]               & 248.353     & [246.976, 249.188]         & 248.11$\pm$0.59        & 248.17$\pm$0.43 \smallskip\\

Mass ratio, $M_c/M_\star$ $^{\bullet}$   &                         & 5.1297$\e{-3}$ & [4.9673, 5.2235]$\e{-3}$ & (5.092$\pm$0.065)$\e{-3}$ & \\
Scaled semi-major axis, $a_c/R_{\star}$              &                         & 207.9       & [177.2, 221.8]             & 200$\pm$12             & \\
${T_0'}_c$$^{\bullet}$\;-\;2\;450\;000                 & [BJD$_{\mathrm{TDB}}$]  & 5486.05     & [5479.20, 5498.80]         & 5488.7$\pm$5.2         & \\
$P'_c$$^{\bullet}$                                   & [d]                     & 673.85      & [671.17, 675.06]           & 673.3$\pm$1.0          & 673.35$\pm$0.84\\
$K'_c$                                               & [\ms]                   & 147.1       & [111.7, 170.5]             & 140$\pm$15             & 141.7$\pm$1.8 \smallskip\\
$\sqrt{e_c}\cos{\omega_c}$$^{\bullet}$                 &                         & 0.0498      & [0.0168, 0.0671]           & 0.042$\pm$0.013        & \\
$\sqrt{e_c}\sin{\omega_c}$$^{\bullet}$                 &                         & -0.42101    & [-0.42600, -0.41624]       & -0.4213$\pm$0.0026     & \\

Planet mass, $M_c$                      &[\Mjup]                  & 8.50        & [3.28, 12.54]              & 7.4$\pm$2.6            & 7.65$\pm$0.27 \smallskip\\

\emph{\bf Data} \smallskip\\
\Kepler\ long-cadence normalisation factor$^{\bullet}$  &   & 1.00000671 & [0.99998568, 1.00001930] & 1.0000027$\pm$0.0000086 & \\
\Kepler\ short-cadence normalisation factor$^{\bullet}$ &   & 1.00000166 & [0.99998736, 1.00002187] & 1.0000046$\pm$0.0000089 & \\
\Kepler\ long-cadence jitter$^{\bullet}$  &        & 1.0363   & [0.9524, 1.1392]     & 1.045$\pm$0.048    & \\
\Kepler\ short-cadence jitter$^{\bullet}$ &        & 1.00137  & [0.98934, 1.02462]   & 1.0073$\pm$0.0089  & \\
HIRES jitter$^{\bullet}$                  &        & 4.07     & [3.33, 6.74]         & 4.7$\pm$1.0        & \\
SOPHIE jitter$^{\bullet}$                 &        & 1.485    & [1.276, 1.987]       & 1.61$\pm$0.20      & \\
HIRES offset$^{\bullet}$                  & [\kms] & 0.0314   & [0.0005, 0.0576]     & 0.029$\pm$0.014    & \\
SOPHIE offset$^{\bullet}$                 & [\kms] & 25.87572 & [25.86130, 25.88755] & 25.8741$\pm$0.0065 & \smallskip\\

\hline

\end{tabular}
\begin{list}{}{}
\item {\bf{Notes.}}
  $^{(\bullet)}$ \emcee\ jump parameter.
  $^{(\dagger)}$ \citet{kipping2013} parametrisation for the limb-darkening coefficients to consider only physical values.
  $^{(\ast)}$ fixed at $t_{\mathrm{ref}}$.\\
  $T'_0 \equiv t_{\mathrm{ref}} - \frac{P'}{2\pi}\left(M_0-E+e\sin{E}\right)$ with $E=2\arctan{\left\{\sqrt{\frac{1-e}{1+e}}\tan{\left[\frac{1}{2}\left(\frac{\pi}{2}-\omega\right)\right]}\right\}}$, $P' \equiv \sqrt{\frac{4\pi^2a^{3}}{\mathcal G M_{\star}}}$, $K' \equiv \frac{M_p \sin{i}}{M_\star^{2/3}\sqrt{1-e^2}}\left(\frac{2 \pi \mathcal G}{P'}\right)^{1/3}$.\\ 
  CODATA 2014: $\mathcal G$ = 6.674$\;$08\ten[-11]~$\rm{m^3\;kg^{-1}\;s^{-2}}$. IAU 2012: \rm{au} = 149$\;$597$\;$870$\;$700~\rm{m}$\;$. IAU 2015: \Rnom = 6.957\ten[8]~\rm{m}, \GMnom = 1.327$\;$124$\;$4\ten[20]~$\rm{m^3\;s^{-2}}$, $\RJnom$ = 7.149$\;$2\ten[7]~\rm{m}, \GMJnom = 1.266$\;$865$\;$3\ten[17]~$\rm{m^3\;s^{-2}}$. \\
  $\Msun$ = \GMnom/$\mathcal G$, \Mjup = \GMJnom/$\mathcal G$, $k^2$ = \GMnom$\;(86\;400~\rm{s})^2$/$\rm{au}^3$
\end{list}
\end{table*}

\subsection{Secular behaviour}

To explore the behaviour of the system at longer timescales, we performed numerical integrations of the system for 10 kyr after the observations. A random sample of size 10,000 from the posterior distribution of the photodynamical modelling was selected as the starting point for the integration. The results for selected parameters are plotted in Fig.~\ref{fig.LongTermEvolutionJacobi}. The simulations show that the orbits librate around apsidal anti-alignment with an amplitude of $18^o$, while the ascending node longitudes librate around alignment with an amplitude of $5^o$, while the mutual inclination undergoes nutation with an amplitude of $2^o$ about $4^o$ \citep{mardling2007,mardling2010}. Furthermore, we put a strong constraint on the mutual inclination of the planets on the longer timescales as well. Our analysis constrains the mutual inclination of the planets to be smaller than 18.9\degree\ at 99\% confidence level (Fig.~\ref{fig.imut}).

These results are highly suggestive of gentle relaxation at some point in the system's history, with the source of dissipation coming from either the protoplanetary disk or tides in planet~b (or both). This is, however, apparently not in agreement with the misaligned orbit suggested by the data (Sect.~\ref{sect.stellarmodels}).

\subsection{Stellar models}\label{sect.stellarmodels}

\begin{figure}
\includegraphics[width=0.48\textwidth]{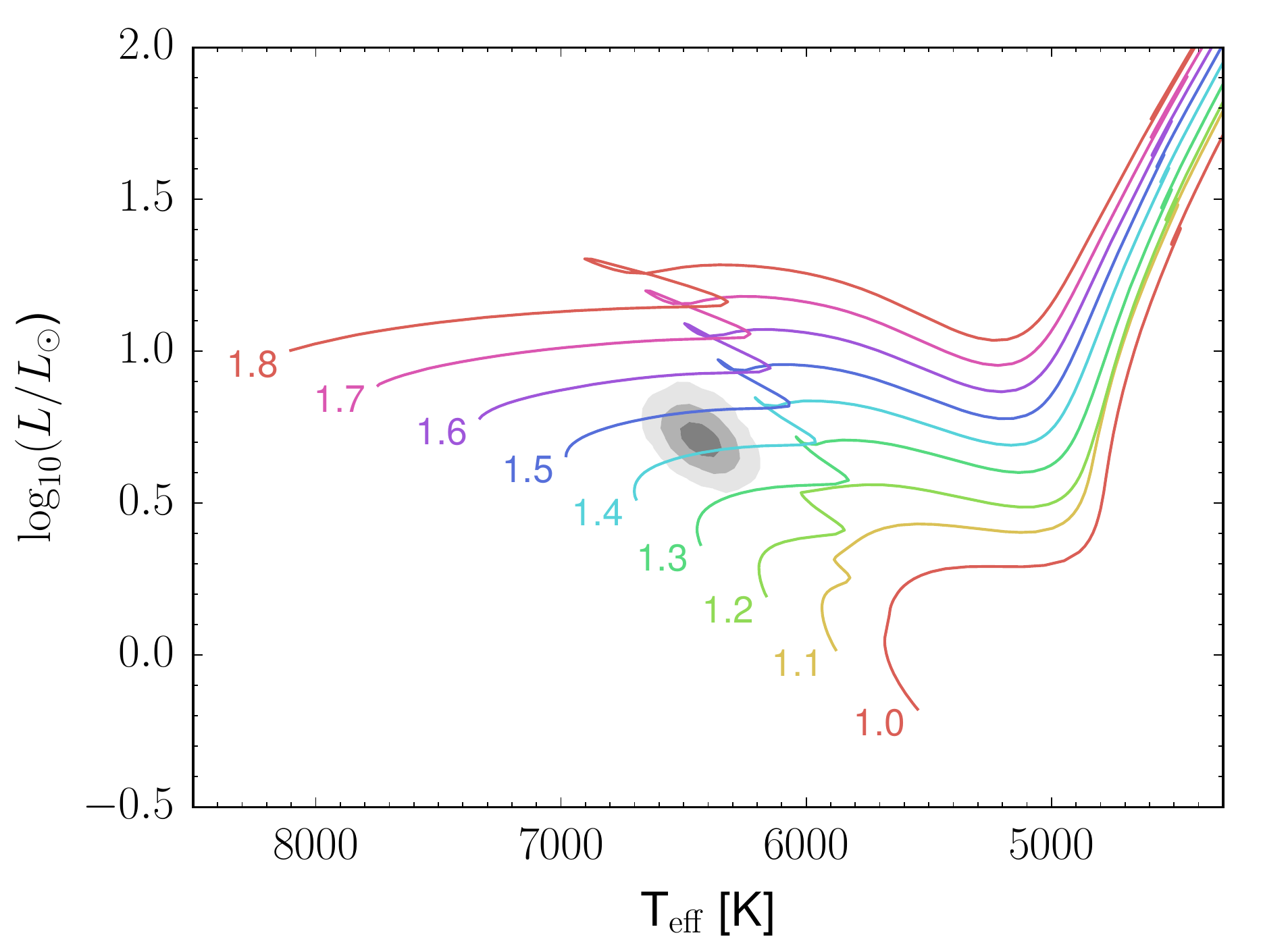}
\caption{Stellar evolution tracks from Dartmouth models, from the zero age main sequence stage to 25 Gyr age, in the luminosity-effective temperature plane for [Fe/H] = 0.15. The results from the photodynamical modelling using stellar models are denoted by the grey contours corresponding to 39.3\%, 86.5\%, and 98.9\% joint confidence regions. The mass in solar masses is annotated at the beginning of the main sequence of each track.}
\label{fig.HR}
\end{figure}

Although we determined absolute masses and radii dynamically, theoretical stellar models can add further constraints on and improve the precision of some parameters. By doing this, stronger assumptions are introduced in the inference process, which might lead to a degraded accuracy. 
We interpolated the Dartmouth models \citep{dotter2008} using the atmospheric parameters from \citet{dawson2014} (T$_{\rm eff}=6430\pm79$~K, [Fe/H]$=0.176\pm0.070$), for which we assumed uncorrelated normal distributions (see Fig.~\ref{fig.HR}), and the stellar density from our model, for which we kept only posterior samples that were compatible with the stellar models. In the last column of Table~\ref{table.results} we list the results for the parameters that were improved significantly by this procedure: the semi-major axes, the masses, and the radii. We note that the new posterior distribution of the stellar radius has less mass outside the limit imposed by the non-detection of p-mode oscillations \citep[$R_\star<1.9~\Rsun,{}$][]{dawson2014} than for the photodynamical determination ($R_\star=1.80\pm0.22~\Rsun{}$).

We obtained an isochronal age of 2.37$\pm$0.31~Gyr, which is about half the stellar main  sequence lifetime for this mass. We compared this age with that derived via gyrochronology. Coupling the rotational period P$_{\rm rot}=4.53\pm0.16$~days \citep{mazeh2015} with the mass determination using stellar models, we derived a gyrochronological age of 2.59$^{+6.4}_{-0.54}$~Gyr \citep{barnes2010,barneskim2010}, where we assumed the zero age main sequence rotational period is  between 0.12 and 3.4~days, and we  added a systematic 10\% error to the statistical error \citep{meibom2015}. Isochronal and gyrochronological ages agree within the uncertainties.

With the observed $\vsini=12.3\pm1.0$~\kms, and the rotational velocity estimated from the rotational period and the stellar radius, it is possible to derive the inclination of the stellar rotational axis $i_\star = 37^{+6}_{-4}$~\degree, in agreement with the more rigorous analysis of \citet{dawson2014}. The orbit of the transiting planet is therefore apparently misaligned with the stellar spin axis. Another star observed to have a coplanar planetary system and a misaligned spin is the red giant Kepler-56 \citep{huber2013}. While most origin scenarios of spin-orbit misalignment involve misalignment of the orbits themselves \citep[e.g.][]{fabrycky2007,nagasawa2008}, both Kepler-419 and Kepler-56 suggest that the truth is more complex.

Finally, the distance to the system was obtained by modelling the spectral energy distribution of Kepler-419 using the PHOENIX/BT-Settl synthetic spectral library \citep{allard2012}, and the procedure described in \citet{diaz2014}. Magnitudes from APASS\footnote{\url{aavso.org/apass}}, 2MASS \citep{2mass}, and WISE \citep{wise} (Table~\ref{table.sed}) were fit to obtain a distance of 993$\pm$67~pc, in agreement with the value obtained by \citet{dawson2012} (see Fig.~\ref{fig.SED}).

\section{Discussion}\label{sec.discussion}

With a first detection reported by \citet{holman2010}, the TTVs method is a recent technique. The \Kepler\ satellite continues to be the only facility to have detected them unambiguously. Confirmations are therefore important in order to validate the method, but also to distinguish possibly degenerated TTV predictions or reveal additional companions. The radial velocity technique is the natural method to do this. However, planets detected both with TTVs and radial velocities are rare today. \citet{barros2014} presented the first radial velocity confirmation of a non-transiting exoplanet discovered by the TTV, and found a mass in agreement with the TTV prediction \citep{nesvorny2013}. In the Kepler-89 system, \citet{masuda2013} predicted from TTVs a mass around 50~\Mearth\ for Kepler-89d, but \citet{weiss2013} measured a mass two times greater using radial velocities. It remains unclear where this difference originates.

Non-transiting planets detected by TTVs are also rare \citep[e.g.][]{nesvorny2012, mancini2016}. Radial velocity detections of such planets would be helpful to confirm and refine the TTV predictions. In cases with degenerate predictions \citep[e.g.][]{ballard2011}, radial velocity measurements would in principle be capable of identifying the correct companion parameters.

In this context, the detection of Kepler-419c in SOPHIE radial velocities is only the second confirmation of a non-transiting exoplanet discovered by TTVs. While \citet{dawson2014} presented the high-precision radial velocity values of this system, they were not sufficient to detect the planet independently, as we did (see Sect.~\ref{sect.keplerian}). The orbital period and velocity amplitude of Kepler-419c measured with SOPHIE agree with the values predicted from the modelling of the photometry alone. This indicates that the model hypotheses are relevant, and that no additional companions  significantly perturb the two detected planets. The dynamical analysis described in Section~\ref{sec.origin} also points to this conclusion. Furthermore, we did not detect the signal of additional planets in the residuals of the radial velocity data.

\begin{figure}
\includegraphics[width=0.48\textwidth]{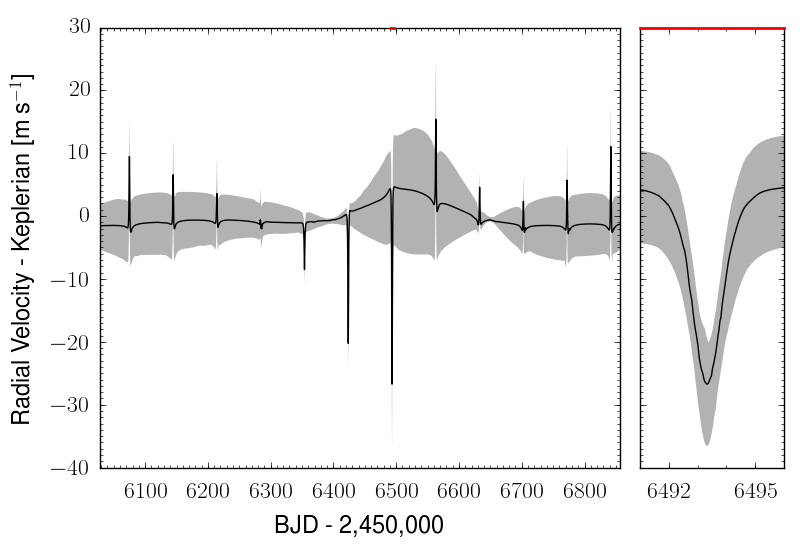}
\caption{Difference between the stellar radial velocity (computed with the n-body integration) and a two-Keplerian fit to the model prediction, computed over 100 random MCMC samples from the full photodynamic model. The black line and the grey shaded region represent the median and the 68.3\% confidence interval, respectively. The smaller panel is an  enlargement around the maximum difference, marked in red in the main panel.
}
\label{fig.diff2Kepl}
\end{figure}

The difference between the radial velocities from the n-body model and those obtained using non-interacting Keplerian curves is relatively small, except close to the periastron passage of planet~b where the difference can be up to $26.7^{+7.3}_{-9.8}$~\ms\ (see Fig.~\ref{fig.diff2Kepl}). The difference increases as the time span of the observations increases. Here we used the combined time span of HIRES and SOPHIE, i.e. 2.3~years.

The main results presented here come from the combined analysis of the \Kepler\ photometry and the SOPHIE radial velocities using a photodynamical model. Under very simple and general assumptions, the model takes into consideration the mutual gravitational interactions of the planets in a consistent manner. A novel result is the dynamic determination of absolute masses and radii. Our results  depend  on the validity of Newtonian mechanics alone, as well as some simple model assumptions: sphericity of all bodies in the system, the number of objects in the system, and the chosen limb-darkening law. To date, Kepler-419 is the system with the most accurate masses and radii determined using the photodynamical model for a multiple planetary system with a single host star\footnote{Circumbinary planetary systems have allowed  much more precise determinations of masses and radii \citep[e.g.][]{doyle2011, orosz2012, kostov2014}.}. While the parameters are determined less precisely than when using theoretical atmosphere and evolutionary stellar models, the limiting factor is certainly the inherent stellar velocity jitter, and to a lesser extent to the photon noise obtained with SOPHIE for this relatively faint star. It is to be expected that more precise results will be achieved for brighter quieter stars, given the same amount of dynamical information. 

Under this model, the radius and mass of planet~b are constrained to 13\% and 24\%, respectively. At the same time, the obtained precision on the stellar mass is 35\%. That is, the mass of planet~b is measured more precisely than the mass of its stellar host. This shows that the constraints provided by our modelling do not imply a measurement relative to the star, as is provided by the Keplerian model of radial velocities. Also, it provides a method for characterising dynamically interacting planets without being limited by knowledge of the star. Ultimately, this technique can be used to test theoretical stellar models, especially for quiet stars. 

\subsection{Mass contraints on Kepler-419 b from photometry}\label{sect.massconstrains}

\citet{dawson2014} affirmed that the TTV of the inner planet are `not at all sensitive' to its mass, and that its mass constraint comes exclusively from the RV data. Our analysis with photometry only (Section~\ref{onlyPH}) shows that an upper limit can be set on the mass ratio of planet~b (Fig.~\ref{fig.mr}). In fact, this non-zero signal is a result of the proximity of the period ratio (which is around 9.7) to 9 and 10, together with the significant eccentricities of both orbits, as we show next.

\begin{figure}
\centering
\includegraphics[width=70mm]{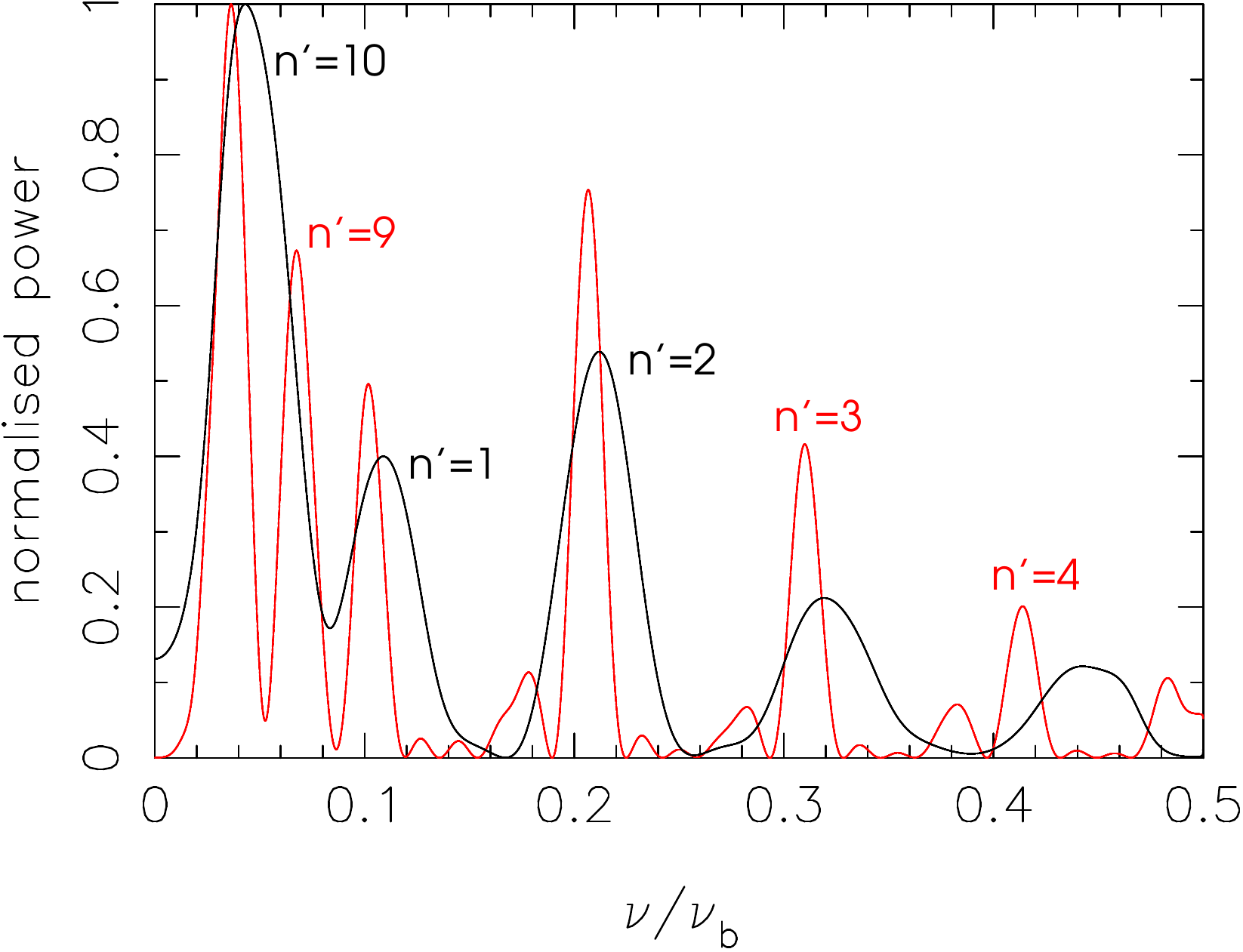}
\caption{Lomb--Scargle periodogram of the TTVs for 22 transits (black curve) and 54 transits (red curve),
with dominant frequencies labelled (see text). Normalised power is plotted against frequency in units of the orbital frequency $\nu_b$ of planet~b. Only the Nyquist window $0\le\nu/\nu_b\le 1/2$ is shown; higher frequencies are aliases of these values.
}
\label{fig.ls}
\end{figure}

Figure~\ref{fig.ls} shows the Lomb--Scargle periodogram (plotted against frequency $\nu$ in units of the inner orbital frequency $\nu_b$) of the 22-transit signal shown in Figure~\ref{fig.TTV}. Only the Nyquist window $0<\nu/\nu_b\le 1/2$ is shown; higher and negative frequencies are aliases of these values and are such that if ${\cal P}(\overline\nu)$ is the power at scaled frequency $\overline\nu\equiv\nu/\nu_b$, then ${\cal P}(\overline\nu)={\cal P}(\overline\nu-k)={\cal P}(k-\overline\nu)$, where $k$ is an integer. \footnote{ These properties are those of the discrete Fourier transform with sampling rate $\nu_b$. Thus, for example, the power at $\overline\nu_2=(\nu_b-2\nu_c)/\nu_b\simeq 0.79$ is the same as the power at $1-\overline\nu_2=0.21$, which is in the Nyquist window. Also, the power at any value of $\nu/\nu_b$ is actually the sum of power in all harmonics with the same value of $n'$ (Mardling, in prep.).} While it is true that the TTV signals of near-circular (single-star) planetary systems are dominated by harmonics associated with first- and second-order resonances (and hence require period ratios of around 3 or less to be detected), the TTVs associated with the 22 transits of Kepler-419b are dominated by harmonics with frequencies $\nu_{n'}\equiv\nu_b-{n'}\nu_c$, $n'=1,2,10$, where $\nu_b$ and $\nu_c$ are the orbital frequencies of planets~b and c, respectively. The substantial power in $n'=10$ is due to both eccentricities (see Table~\ref{sweet}). 

To verify this, we simulated a light curve with 54 transits of planet~b using the MAP values in Table~\ref{table.results} with a sampling and white noise amplitude equal to that of the Kepler SC data. The TTV periodogram for the simulated LC is also shown in Fig.~\ref{fig.ls}. It reveals that the dominant harmonic is $n'=10$, and that the powerful $n'=9$ harmonic is not resolved in the 22-transit data set\footnote{At least 44 transits are required before the $n'=9$ harmonic emerges for these initial conditions.}. This lack of resolution results in a poor constraint on the mass of the transiting planet for the following reasons.

For a coplanar system there are seven unknowns 
(the masses of the two planets, the two eccentricities and corresponding longitudes of periastron, and the mean longitude of planet~c at epoch). Therefore, information from at least four well-resolved harmonics (each with an amplitude and a phase) is needed in order to place reasonable bounds on all seven parameters from TTV data alone. We can show analytically (Mardling, in prep.) that the back effect of the mass of planet~b on its own TTVs is to reduce the amplitude of the $n'$th harmonic by a factor of approximately $1-n'(M_b/M_*)$. Thus, for the low-order harmonics $n'=1,2,3$, typically the most powerful detectable for low-eccentricity systems, the difference is of the order of a fraction of a per cent. On the other hand, for $n'=9$ and $n'=10$ the effect is at the 2\% level, and can therefore be detected, provided they can be correctly resolved.

 \begin{table}
 \caption{L-S harmonic power showing eccentricity sweet spot.}
 \label{sweet}
 \begin{tabular}{cccccccc}
 Case & $e_b$ & $e_c$ & ${\cal P}_{10}$ & ${\cal P}_9$ &  ${\cal P}_1$ &  ${\cal P}_2$ &  ${\cal P}_3$  \\ \hline
$1$ & 0.8070 & 0.1797 & {\bf 1.00} & 0.67 & 0.50 & 0.75 & 0.42  \\
$2$ & 0.8070 & 0.1 & 0.04 & 0.05 & 0.39 & {\bf 1.00} & 0.35  \\
$3$ & 0.8070 & 0.3 & {\bf 1.00} & 0.37 & 0.03 & 0.01 & 0.01  \\
$4$ & 0.5 & 0.1797 & 0.16 & 0.15 & 0.44 & {\bf 1.00}  & 0.51  \\
$5$ & 0.9 & 0.1797 & {\bf 1.00}  & 0.59 & 0.39 & 0.49 & 0.27 \\
\hline
 \end{tabular}
 \end{table}
Table~\ref{sweet} lists the normalised Lomb--Scargle power, ${\cal P}_{n'}$, for $n'=\{10, 9, 1, 2, 3\}$, for values of the inner and outer eccentricities around the MAP values for 54 transits. The dominant harmonic is highlighted in bold. We note that substantial power is associated with all five harmonics in the observed system (case 1), in contrast with cases 2 and 3 for which the eccentricity of planet~c is respectively reduced and increased  by around 0.1. Similarly, changing the eccentricity of the transiting planet significantly affects the distribution of power in the various harmonics. Thus, we see that the Kepler-419 system is in a serendipitous sweet spot for the determination of system parameters via TTVs (given a sufficient number of transits). This is confirmed by the filled grey histograms in Figure~\ref{fig.mr}, showing the posterior sample the masses of planets~b (left) and c (right), based on the photodynamical analysis of the simulated light curve with 54 transits (and no radial velocities). Not only is the mass of the transiting planet well resolved, the mass of the perturbing planet is also  significantly better resolved than it is with just 22 transits, with or without radial velocities.

\subsection{Origin of the Kepler-419 system}\label{sec.origin}

The coplanar yet highly eccentric nature of Kepler-419 presents a puzzle regarding its origin. The system's proximity to a stable fixed-point suggests that it is likely to have been brought gently to this relaxed state via disk dissipation or planet (or even stellar) tides or both.

\begin{figure}
\centering
\includegraphics[width=60mm]{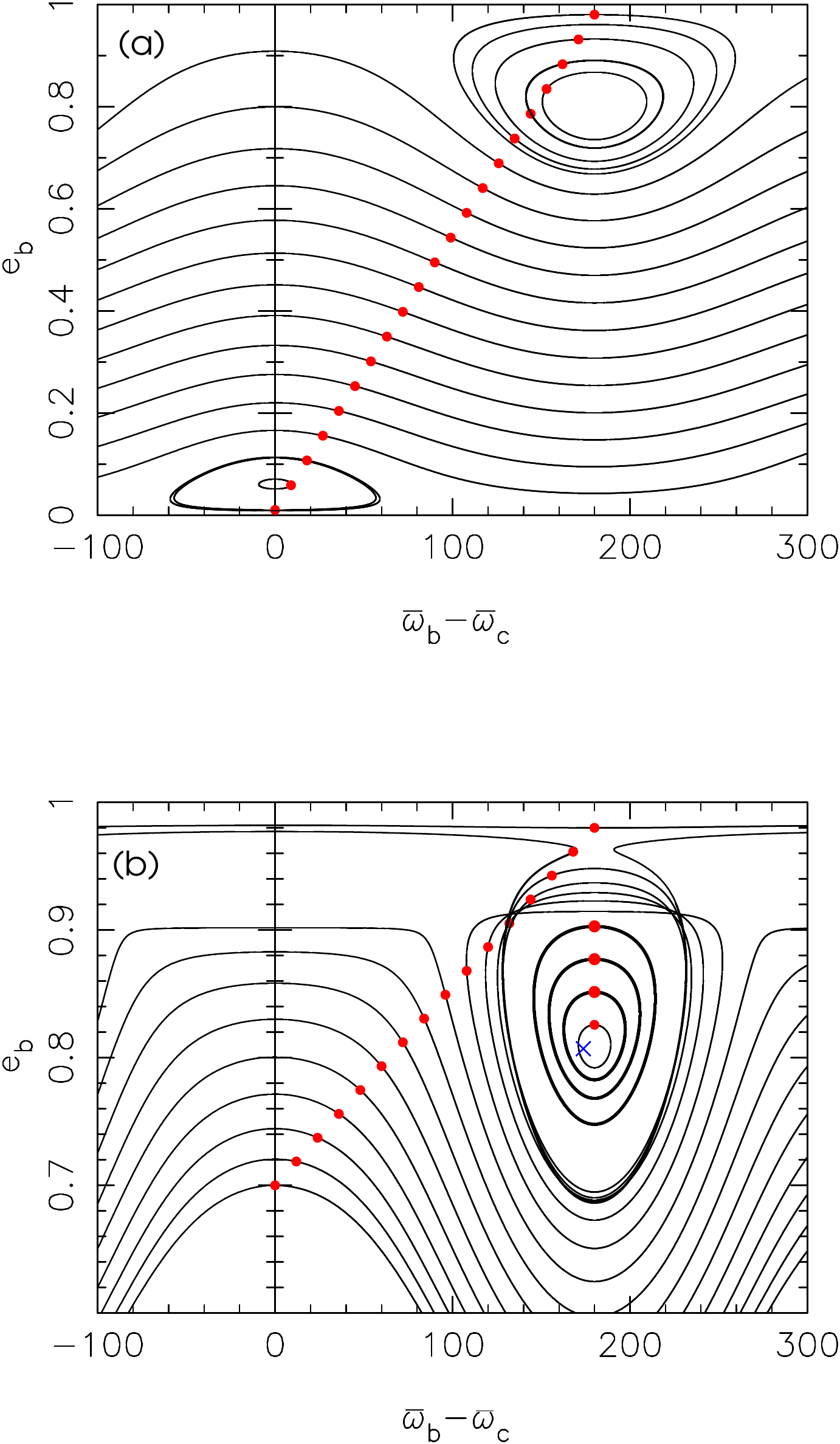}
\caption{Variation of $e_b$ as a function of $\varpi_b-\varpi_c$ in the absence of damping, 
showing the existence of low- and high-eccentricity fixed points. Curves produced by integrating
the secular equations of motion with the Kepler-419 system parameters as initial conditions except for
$e_b$ and $\varpi_b-\varpi_c$, which are indicated by red dots.
In panel (a), $e_c$ is held constant, while in panel (b), $e_c$
is allowed to vary. Phase curves which cross have different values of $e_c$ at those points. 
The blue cross shows the position of Kepler-419
and its proximity to the fixed point.
We note the different $y$-axis scales.
}
\label{fig.FP}
\end{figure} 
While the existence of low-eccentricity fixed points in coplanar systems is well known  \citep[e.g.][]{mardling2007}, as is the existence of high-eccentricity fixed points in non-coplanar systems \citep[those associated with the Kozai mechanism; e.g.][]{naoz2016}, the existence of high-eccentricity fixed points in coplanar systems \citep[e.g.][]{nagasawa2003} seems to be less well known. Figure~\ref{fig.FP} shows the existence of low- and high-eccentricity fixed points, with curves produced by integrating the secular equations of motion (in the absence of damping), with the  Kepler-419 system parameters as initial conditions except for $e_b$ and $\varpi_b-\varpi_c$, which are indicated by red dots.
In panel (a) $e_c$ is held constant, while in panel (b) $e_c$ is allowed to vary. 
The blue cross shows the position of Kepler-419 and its proximity to the fixed point, suggesting that it is highly likely that the system has relaxed towards that state.

In order to reach planet~b’s high eccentricity, a significant source of external torque must be identified, which raises the eccentricity and also maintains the coplanarity of the planets. In the following we discuss the viability of various mechanisms capable of providing such a torque.

\subsubsection*{External forcing}
\citet{dawson2014} considered Kozai forcing of planet~b by planet~c, but dismissed it because the mutual inclination (here constrained even further to be no more than a few degrees) is not compatible with the large value normally necessary to achieve eccentricity growth. A potential solution to this quandary is that a significantly inclined external fourth body has brought the system to its current relaxed state.
\citet{takeda2008} showed that under favourable circumstances a pair of planets will respond to the torque from such a body and will remain almost coplanar while undergoing Kozai oscillations. 
The key to this mechanism is that the resulting precession rates of each planetary orbit should be similar so that their response is in concert, with the difference in the node angles librating around a fixed value.
However, this cannot be the case for the Kepler-419 system because the mutual torques are already almost perfectly balanced; the addition of a fourth body would destroy this harmony. 

\citet{dawson2014} also dismissed the `orbit flip' mechanism \citep{li2014} because they did not observe this to happen in integrations using their observed orbital parameters. 
This mechanism initially involves  almost coplanar orbits periodically undergoing large excursions in relative inclination and (inner) eccentricity, with periodic $180^o$ flips in the orientation of the orbital angular momentum vector of a body relative to that of an external perturber. We note that the  \citet{li2014} analysis is done in the test particle approximation and therefore does not involve any variations in the perturber's orbit.
It is associated with octopole-level terms in the disturbing function\footnote{The usual Kozai process is governed by quadrupole terms.}, and occurs for favourable orbital configurations which satisfy an analytic condition involving the relative strengths of the octopole and quadrupole terms.
In fact, we can study the problem from the point of view of finding the fixed points of the octopole-level equations of motion for the general problem (all three bodies massive) and enquire about their stability (Mardling, in prep.). 
We find that flip-type solutions are associated with unstable fixed points, and in the presence of damping a system would evolve away from such a state.
It therefore seems unlikely that a real system would be observed in a `flip' configuration, and it is clear that the Kepler-419 system is not in this state.

\subsubsection*{Spin-orbit coupling}
Another promising mechanism is spin-orbit coupling between the planet and effectively {both} orbits.
\citet{correia2012} showed that under favourable circumstances, the eccentricity of the orbit of a spinning planet in a two-planet system can be raised to high values.
This mechanism relies on the fact that the spin-synchronisation timescale of a planet is much shorter than the eccentricity-damping timescale, and so is effective for relatively long-period orbits.
Since a companion will modulate the eccentricity of the tidally active planet on the secular timescale, and since the spin will tend to synchronise with the orbital frequency at periastron, but will lag behind the eccentricity forcing by an amount that depends on the spin-synchronisation timescale, a small positive drift in the average eccentricity can result if the two timescales are similar. 
The average eccentricity continues to increase until torques are balanced (the system reaches a fixed point).
However, tidal heating at periastron also increases (because secular forcing does not change the  semi-major axis and hence decreases the periastron separation as the eccentricity increases), ultimately shrinking the semi-major axis and circularizing the orbit on a timescale which may be longer than the age of the system \citep[see Fig.~1 in][]{correia2012}.
Using a double-averaged code with spin-orbit coupling, relativistic apsidal advance and tides \citep{mardling2002}, with realistic structure and damping parameters and the current observed parameters of the Kepler-419 system but with low initial values of the inner eccentricity and arbitrary initial values of the difference in the longitudes of periastron $\varpi_b-\varpi_c$, some positive drift in the latter is observed for some initial configurations but appears to be quite sensitive to the initial value of $\varpi_b-\varpi_c$. 
None of the systems considered experienced eccentricity increases of much more than 0.1 on the several Gyr timescale; however, if  some gentle disk migration  (weak enough to avoid resonance capture) is included, thereby providing additional torques, as well as the evolution of the planet radius due to gravitational contraction (resulting in stronger tides in the past), it is conceivable that the mechanism could produce the Kepler-419 system as it is currently observed.

\subsubsection*{Collision and damping}
Finally we consider a possible collision and subsequent damping scenario for the origin of Kepler-419, and examine the nature of the high-eccentricity fixed point and its implications for the existence of hot Jupiters.

Consider the scenario in which the Kepler-419 system forms a three-planet $1\!:\!2\!:\!4$ Laplace configuration via inward convergent migration of three (or initially more) planets, with the current planet~c on the outside. 
Notwithstanding the protective nature of the Laplace resonance, eccentricity growth can result in orbit crossing and eventual collision of the inner two planets for favourable systems and disk conditions.
As far as we are aware, fully self-consistent modelling of such an event has not been done, and as such it remains unclear how angular momentum would be distributed between the resultant planet orbits and the ejecta following collision.
It is conceivable that the post-collision orbit could have considerably less angular momentum than that of the precursor orbits, with the excess being returned to the protoplanetary disk or escaping the system.

A less constraining scenario is for the pre-cursor system to consist of a pair of planets in a 2:1 resonance interior to the current planet~c (but not participating in a Laplace-type configuration), which again undergo collision following eccentricity pumping and subsequent instability \citep{goldreich2014}. 
Either way, we propose a scenario where a collision results in a highly eccentric orbit whose apsidal orientation with respect to that of planet~c is such that it is immediately or subsequently captured by the high-eccentricity fixed point associated with coplanar two-planet systems.

Depending on the initial value of the angle between the apsidal lines, such a state can in fact prevent a highly eccentric system from circularising, as appears to be the case for Kepler-419. Alternatively, it can result in a hot Jupiter, as the following demonstrates.

\subsubsection*{Formation of hot Jupiters}
\begin{figure*}
\centering
\includegraphics[width=180mm]{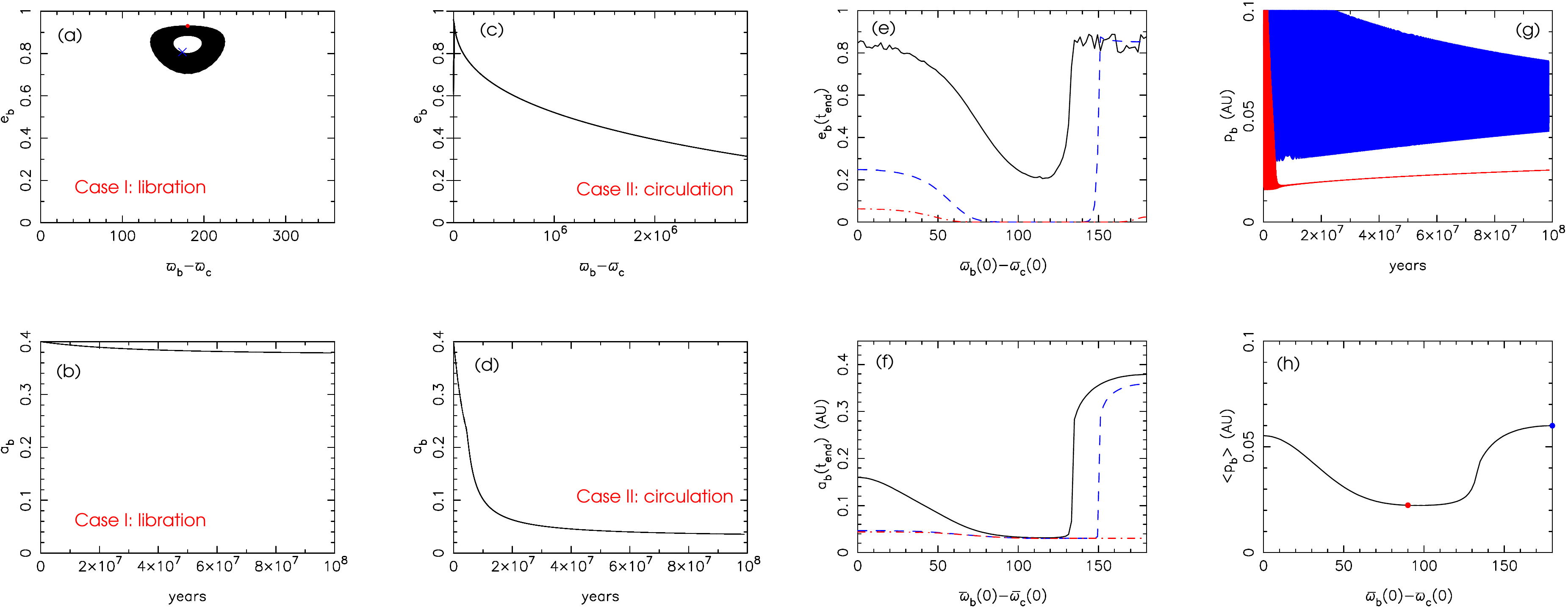}
\caption{Effect of the initial value of $\varpi_b-\varpi_c$ on the subsequent tidal evolution of a system. Panels (a) and (b): $\varpi_b(0)-\varpi_c(0)=180^o$ (Case I); panels (c) and (d): $\varpi_b(0)-\varpi_c(0)=90^o$ (Case II).
The red dot and blue cross in panel (a) indicate the initial and current state, respectively, of Kepler-419. Panels (e) and (f): Values of $a_b$ and $e_b$ at $t_{end}=10^8$ yr (solid black curves), $t_{end}=10^9$ yr (dashed blue curves), and $t_{end}=2.3\times10^9$ yr (the estimated age of the system; dot-dashed red curves) as functions of the initial value of $\varpi_b-\varpi_c$, which show that for a range of initial relative orientations, planet~b can sustain a high value of $e_b$ because it spends most of its time with a periastron distance which is too high for significant circularisation to occur. Panel (g): Evolution of planet~b's periastron distance, $p_b$, for Case I (blue) and Case II (red), the latter showing escape from libration and hence a permanently low value of $p_b$. Panel (h): Average value of $p_b$, $\langle p_b\rangle$, as a function of $\varpi_b(0)-\varpi_c(0)$. Angles are in degrees.}
\label{fig.evolve}
\end{figure*}

When damping due to planetary tides is included, the semi-major axis of planet~b responds by shrinking at a rate which depends on the minimum value of the periastron separation over a secular cycle and the time spent near that value.
Panels (a)--(d) of Figure~\ref{fig.evolve} show the evolution of the eccentricity and semi-major axis for two systems which are identical except for the initial value of $\varpi_b-\varpi_c$. For panels (a) and (b) (case I), $\varpi_b(0)-\varpi_c(0)=180^o$, while $\varpi_b(0)-\varpi_c(0)=90^o$ for panels (c) and (d) (case II). The remaining system parameters are as for Kepler-419, except that $a_b(0)=0.4$ and $e_b(0)=0.93$. The $Q$-value and Love number of the planet are taken as $10^5$ and 0.3, respectively. The first $10^8$~years of evolution is shown.
Very little orbit shrinkage and circularisation has occurred for case I after $10^8$~years, with the system remaining trapped near the fixed point at $(\varpi_b-\varpi_c=\pi,e_b=0.83)$ with an average periastron separation of 0.06~au.
In contrast, case II initially achieves a maximum eccentricity of 0.96 during a short period of libration, after which it escapes the librating region with a non-oscillatory eccentricity, and thus a permanently low periastron separation. With an average periastron separation of 0.02~au, circularisation is rapid. We emphasise that the only difference between the two cases is the inital value of the difference in the apsidal longitudes.

Panels (e) and (f) show the dependence of the values of $e_b$ and $a_b$ at $10^8$~yr, $10^9$~yr, and $2.3\times 10^9$~yr (the estimated age of the system) on the initial difference in the apsidal longitudes, suggesting that the long-term state of a system like Kepler-419 is highly dependent on the value of $\varpi_b-\varpi_c$ at the time it was brought to that state (via collision or some other mechanism).
For the adopted $Q$-value the system has circularised after $2.3\times 10^9$~yr, suggesting that the true $Q$-value is higher or has varied over the lifetime of the system, or that the system is younger, or both.
Panel (g) shows the evolution of the periastron distance of planet~b for Cases I and II, while panel (h) shows its average value, $\langle p_b\rangle$, as a function of the initial difference in the apsidal longitudes. Thus while a system is trapped in the libration state it tends to maintain a high eccentricity because $\langle p_b\rangle$ is relatively high, while escape from libration is associated with a permanently low value of $p_b$.

This suggests that a system like HAT-P-13 \citep{bakos2009}, whose periods and planet masses are 2.9~d and 428.5~d, and 0.85~\Mjup\ and 15.2~\Mjup, respectively, and whose star has a mass of $1.22~M_\odot$, may have had a similar origin to Kepler-419, but found itself with a value of $\varpi_b-\varpi_c$ conducive to circularisation. 
Just as for the Kozai migration mechanism, such an evolutionary history would leave little room for intervening planets. 
However,  unlike the Kozai mechanism, the present mechanism produces hot Jupiters without the need of mutual inclination.

%


\begin{acknowledgements}
    We thank Rebekah I. Dawson and Geoffrey W. Marcy for helpful discussions.   
  We thank the technical team at the Observatoire de Haute-Provence for their support with the SOPHIE instrument and the 1.93m telescope and, in particular, for the essential work of the night assistants. Financial support for the SOPHIE observations from the Programme National de Plan\`etologie (PNP) of CNRS/INSU, France, is gratefully acknowledged. We also acknowledge support from the French National Research Agency (ANR-08- JCJC-0102-01). We thank L. Kreidberg for her Mandel \& Agol code and Y. Revaz for his assistance with the computing cluster used in this work. This paper includes data collected by the \Kepler\ mission. Funding for the \Kepler\ mission is provided by the NASA Science Mission directorate. Some of the data presented in this paper were obtained from the Mikulski Archive for Space Telescopes (MAST). This research has made use of the Exoplanet Orbit Database and the Exoplanet Data Explorer at exoplanets.org.
  This research was made possible through the use of the AAVSO Photometric All-Sky Survey (APASS), funded by the Robert Martin Ayers Sciences Fund.
  This publication makes use of data products from the Two Micron All Sky Survey, which is a joint project of the University of Massachusetts and the Infrared Processing and Analysis Center/California Institute of Technology, funded by the National Aeronautics and Space Administration and the National Science Foundation.
  This publication makes use of data products from the Wide-field Infrared Survey Explorer, which is a joint project of the University of California, Los Angeles, and the Jet Propulsion Laboratory/California Institute of Technology, funded by the National Aeronautics and Space Administration.
  Simulations in this paper made use of the \reb\ code which can be downloaded freely at \texttt{http://github.com/hannorein/rebound}. These simulations have been run on the {\it Regor} cluster kindly provided by the Observatoire de Gen\`eve. This publication makes use of The Data \& Analysis Center for Exoplanets (DACE), which is a facility based at the University of Geneva (CH) dedicated to extrasolar planets data visualisation, exchange, and analysis. DACE is a platform of the Swiss National Centre of Competence in Research (NCCR) PlanetS, federating the Swiss expertise in Exoplanet research. The DACE platform is available at \texttt{https://dace.unige.ch}. This work has been carried out within the framework of the National Centre for Competence in Research PlanetS supported by the Swiss National Science Foundation. The authors acknowledge the financial support of the SNSF. S.C.C.B. acknowledges support by the Funda\c c\~ao para a Ci\^encia e a Tecnologia (FCT) through the Investigador FCT Contract IF/01312/2014/CP1215/CT0004 and  also acknowledges support from FCT through national funds and by FEDER through COMPETE2020 by these grants UID/FIS/04434/2013 \& POCI-01-0145-FEDER-007672 and PTDC/FIS-AST/1526/2014 \& POCI-01-0145-FEDER-016886.
  A.S.B. acknowledges funding from the European Union Seventh Framework programme (FP7/2007-2013) under grant agreement No. 313014 (ETAEARTH).

\end{acknowledgements}

\bibliographystyle{aa}
\bibliography{k419}

\begin{appendix}

\section{Additional figures and tables}

\begin{table}
\small
  \caption{SOPHIE radial velocity measurements of Kepler-419.}
\begin{tabular}{lccc}
\hline
\hline
BJD - 2\,400\,000 & RV & $\pm1\sigma$ \\
& [km\,s$^{-1}$] & [km\,s$^{-1}$]  \\
\hline
56064.49701 & -25.671 & 0.039 \\
56133.47769 & -25.792 & 0.020 \\
56155.58275 & -25.954 & 0.031 \\
56159.48998 & -25.915 & 0.023 \\
56163.55945 & -25.950 & 0.028 \\
56181.49327 & -25.863 & 0.023 \\
56213.43735 & -25.752 & 0.029 \\
56271.24436 & -25.927 & 0.032 \\
56378.65431 & -26.158 & 0.038 \\
56401.59330 & -26.005 & 0.021 \\
56416.54230 & -25.769 & 0.024 \\
56449.51760 & -25.986 & 0.022 \\
56472.46928 & -25.839 & 0.037 \\
56478.47969 & -25.904 & 0.020 \\
56479.49468 & -25.814 & 0.018 \\
56484.41333 & -25.855 & 0.028 \\
56503.54966 & -25.951 & 0.029 \\
56504.52153 & -25.983 & 0.022 \\
56505.47037 & -25.941 & 0.019 \\
56506.39579 & -25.977 & 0.019 \\
56508.43157 & -25.846 & 0.023 \\
56509.50194 & -25.995 & 0.024 \\
56510.49200 & -25.893 & 0.026 \\
56513.57888 & -25.926 & 0.029 \\
56514.51141 & -25.959 & 0.037 \\
56515.51591 & -25.853 & 0.026 \\
56516.57042 & -25.909 & 0.025 \\
56535.41082 & -25.790 & 0.019 \\
56552.39025 & -25.728 & 0.045 \\
56567.36208 & -25.926 & 0.022 \\
56599.31586 & -25.724 & 0.036 \\
56621.31826 & -25.624 & 0.024 \\
56725.62306 & -25.833 & 0.029 \\
56771.58947 & -25.678 & 0.025 \\
56772.59203 & -25.844 & 0.058 \\
56773.60591 & -25.974 & 0.028 \\
56775.57499 & -25.996 & 0.040 \\
56776.60246 & -25.925 & 0.034 \\
56807.51961 & -25.860 & 0.028 \\
56829.54800 & -25.769 & 0.027 \\
56836.53369 & -25.769 & 0.027 \\
56841.45968 & -25.737 & 0.031 \\
56844.47533 & -26.089 & 0.028 \\
56849.56355 & -25.977 & 0.036 \\
56856.53192 & -25.979 & 0.029 \\
\hline
\end{tabular}
\label{table.RV}
\end{table}

\begin{figure*}
\hspace{-2cm}\includegraphics[height=22cm]{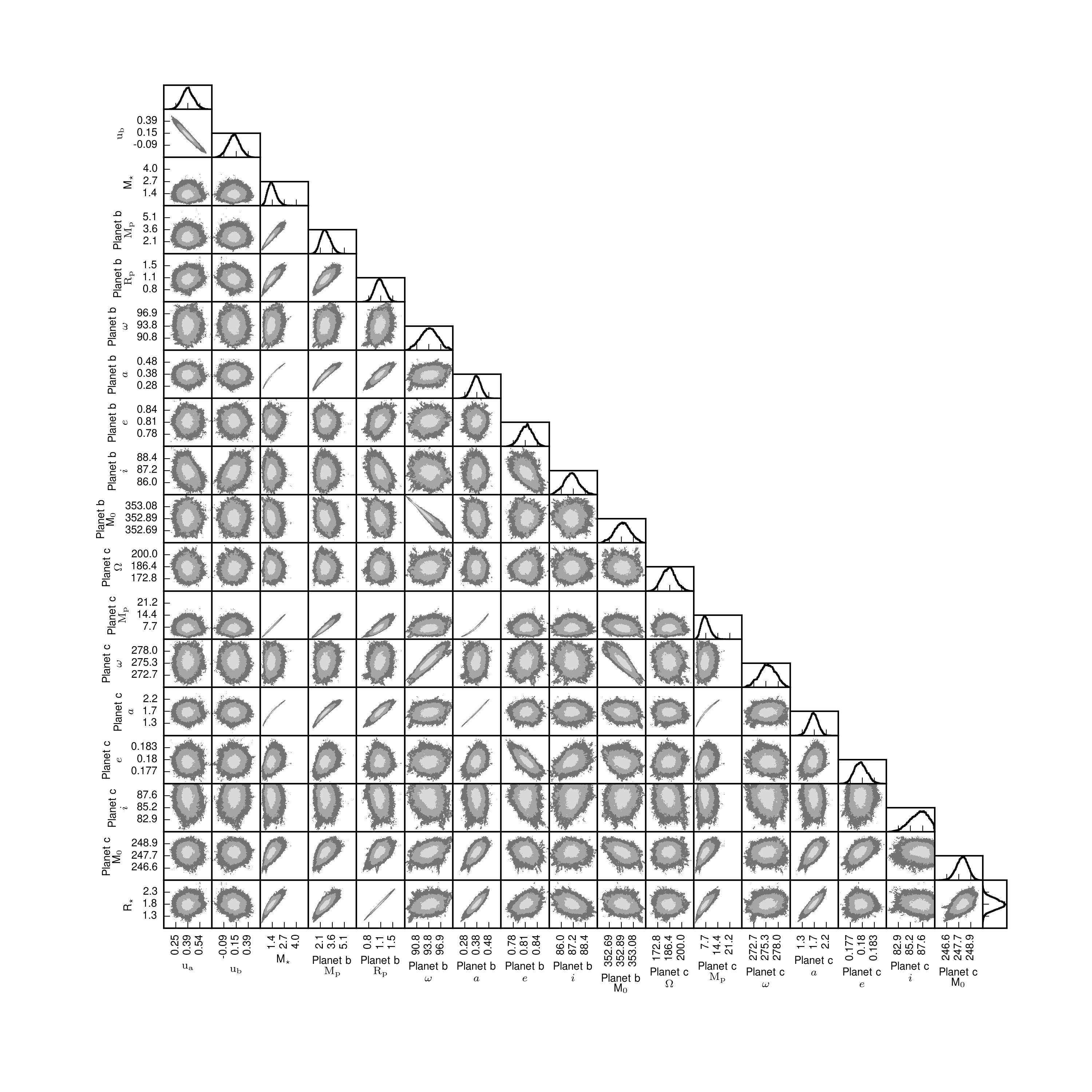}
\vspace{-2cm}\caption{Two-parameter joint posterior distributions for the most relevant MCMC model parameters. The 39.3, 86.5, and 98.9\% two-variable joint confidence regions are denoted by three different grey levels;  in the case of a Gaussian posterior, these regions project on to the one-dimensional 1, 2, and 3~$\sigma$ intervals. The histogram of the marginal distribution for each parameter is shown at the top of each column, except for the parameter on the last line, which is shown at the end of the line. Units are the same as in Table~\ref{table.results}.}
\label{fig.pyramid}
\end{figure*}

\begin{figure*}
\centering
\includegraphics[height=5.1cm]{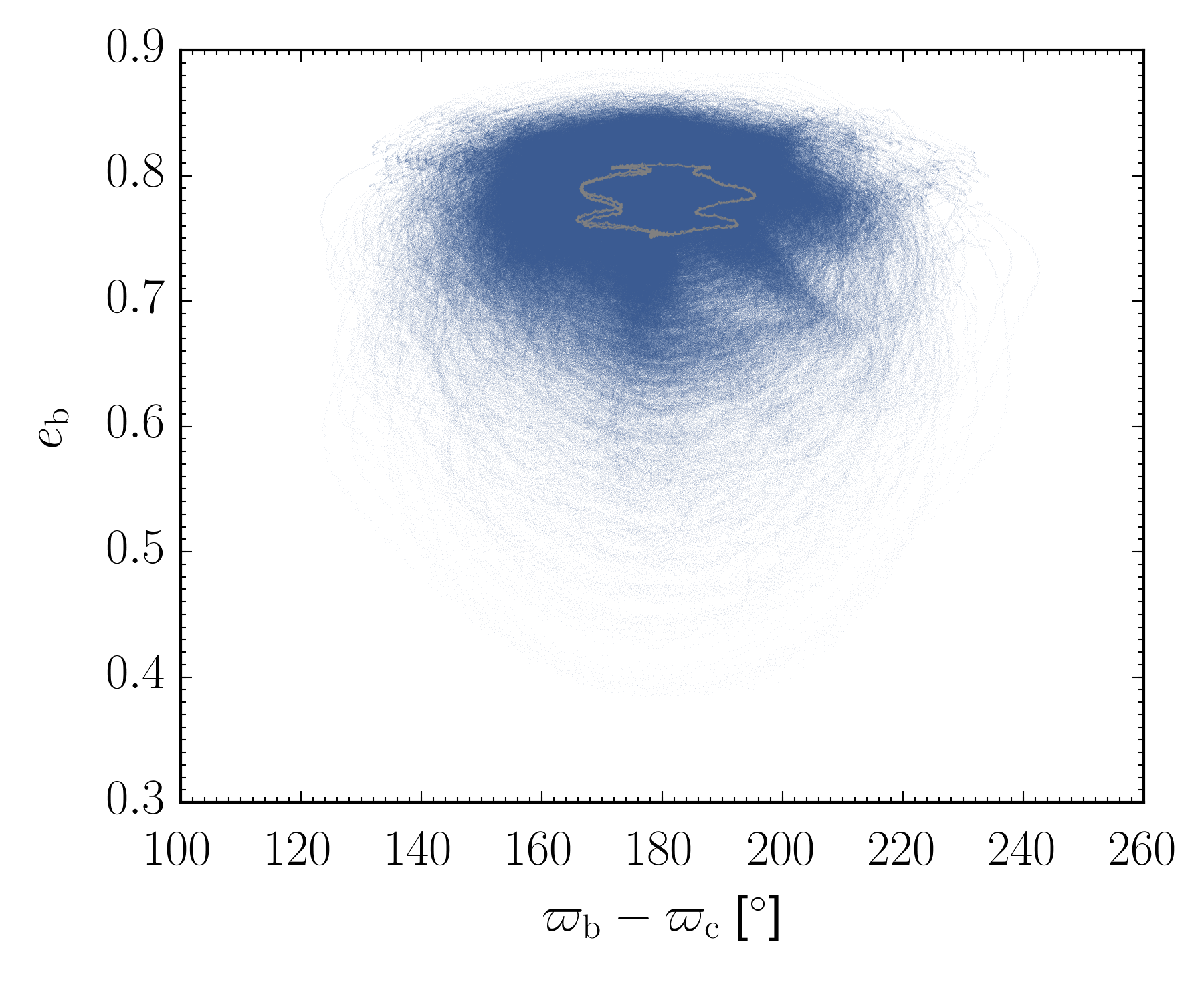}\includegraphics[height=5.1cm]{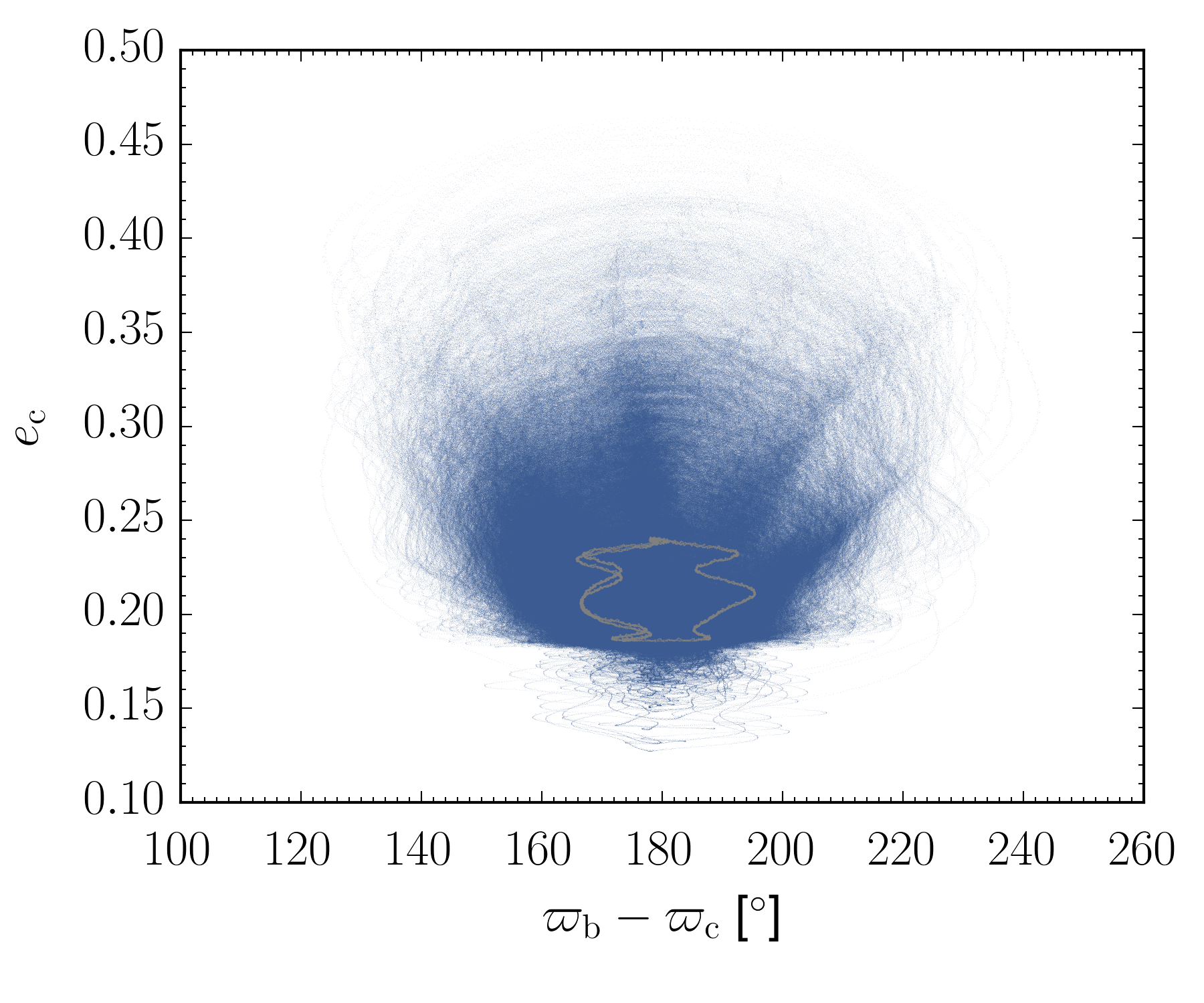}\includegraphics[height=5.1cm]{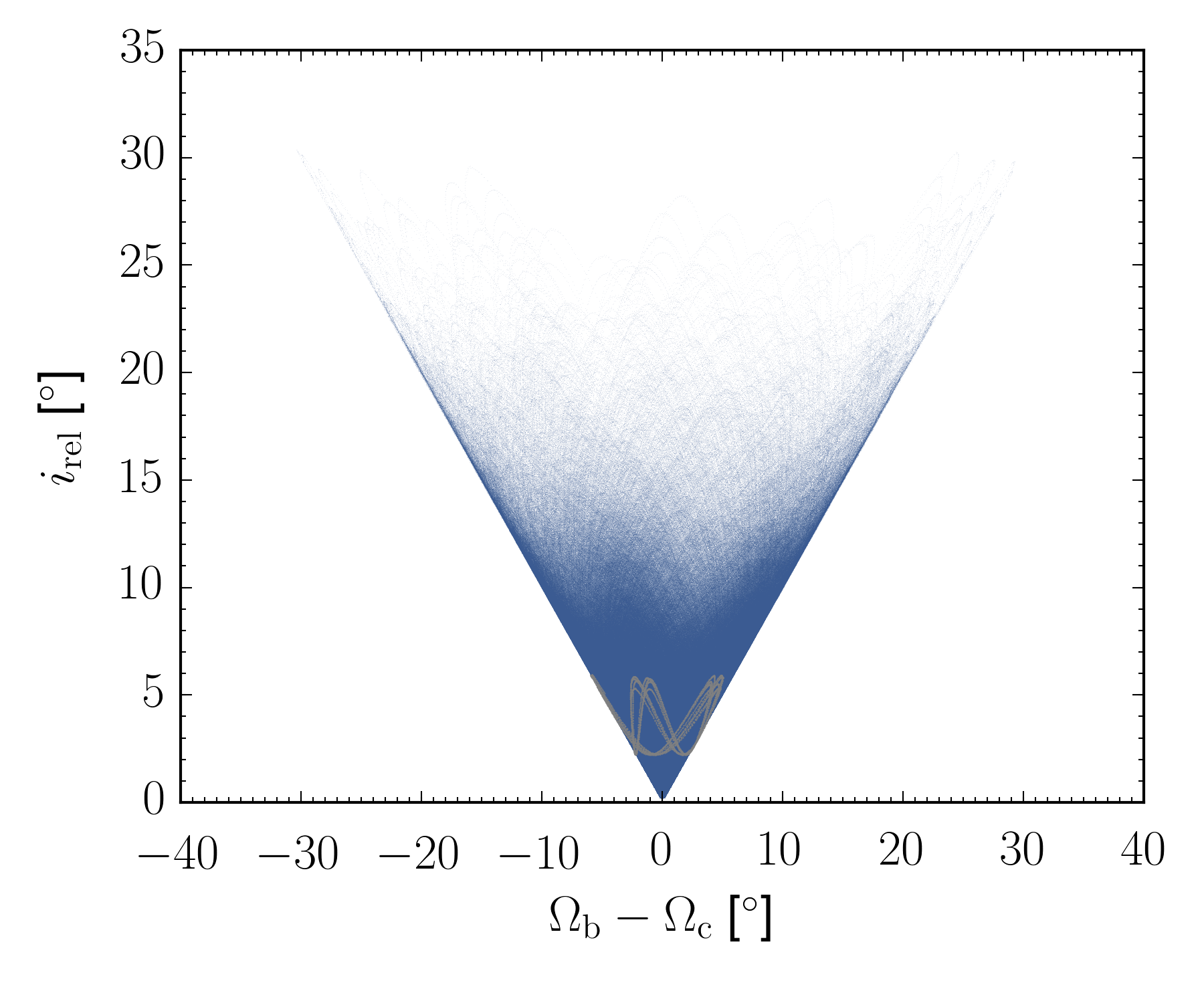}
\caption{Evolution of the Jacobi orbital elements ($\varpi=\Omega+\omega$, is the longitude of the periapsis) over 10 kyr since the beginning of \Kepler\ observations. 10,000 random draws from the posterior distribution are shown. The grey points correspond to the integration using the MAP values.}
\label{fig.LongTermEvolutionJacobi}
\end{figure*}

\begin{figure*}
\centering
\includegraphics[width=0.48\textwidth]{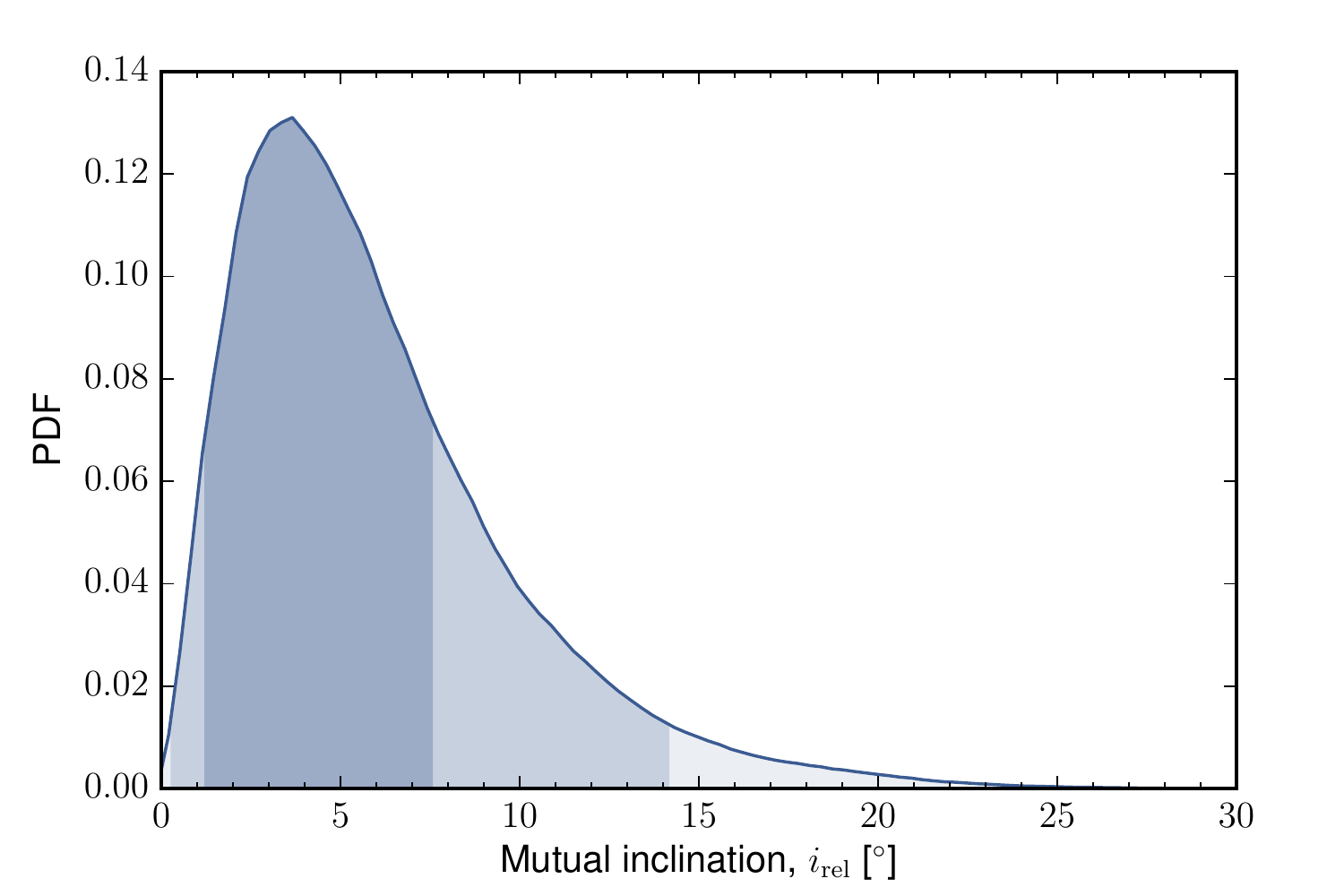}
\caption{Posterior probability function of the mutual inclination of the planets in Jacobi coordinates, based on integrations over 10 kyr. The shaded regions correspond to the 68.3\% [1.20, 7.58], 95.5\% [0.26, 14.17], and 99.7\% [0.05, 22.02] highest density intervals.}
\label{fig.imut}
\end{figure*}

\begin{figure*}
\centering
\includegraphics[width=0.48\textwidth]{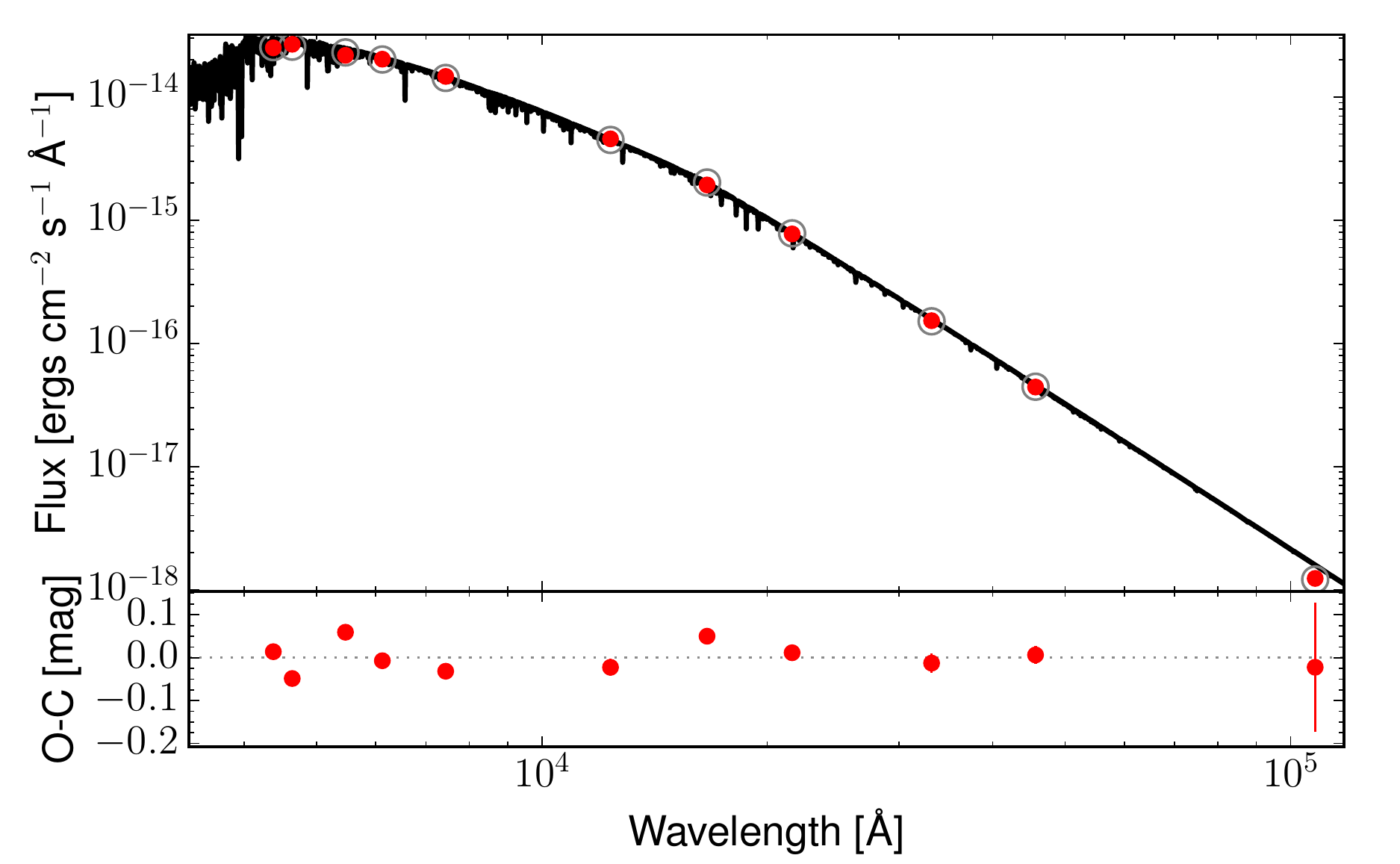}
\caption{Spectral energy distribution of Kepler-419 based on magnitudes from the literature (red circles; see text for details). The best-fit spectrum is plotted as a solid black curve, and the integrated fluxes in the photometric bands are plotted as open circles. The residuals are given in the bottom panel.}
\label{fig.SED}
\end{figure*}

\begin{table}
\small
  \caption{Photometric measurements used for the SED of Kepler-419.}
\begin{tabular}{lccl}
\hline
\hline
Filter & Magnitude & $\pm1\sigma$ & Source \\
\hline
Johnson-B  & 13.498  &  0.011 & APASS DR9\\
Johnson-V  & 13.036  &  0.006 & APASS DR9\\
SDSS-G     & 13.188  &  0.010 & APASS DR9\\
SDSS-R     & 12.888  &  0.010 & APASS DR9\\
SDSS-I     & 12.815  &  0.006 & APASS DR9\\
2MASS-J    & 12.088  &  0.020 & 2MASS\\
2MASS-H    & 11.899  &  0.019 & 2MASS\\
2MASS-Ks   & 11.859  &  0.018 & 2MASS\\
WISE-W1    & 11.829  &  0.023 & WISE\\
WISE-W2    & 11.851  &  0.021 & WISE\\
WISE-W3    & 11.815  &  0.151 & WISE\\
\hline
\end{tabular}
\label{table.sed}
\end{table}

\end{appendix}
\end{document}